\pdfoutput=1
\documentclass[a4paper,12pt,english]{article}

\usepackage[bindingoffset=0.3cm,textheight=22.5cm,hdivide={2.7cm,*,2.7cm}, vdivide={*,22cm,*}]{geometry}
\usepackage{cite}
\usepackage{youngtab}
\usepackage{amsmath,amsfonts,amssymb,babel,slashed,color,empheq,tensor,amsthm}
\usepackage{tensor}
\usepackage{hyperref}

\makeatletter

\newcommand{\del}{\partial}

\def\nn{\nonumber} 

\numberwithin{equation}{section}

\def\a{\alpha}  \def\b{\beta}
 \def\g{\gamma} 
 \def\d{\delta} 
    \def\k{\kappa}
 \def\L{\Lambda}  
\def\n{\nu}    \def\r{\rho}
\def\s{\sigma}

\def\cA{{\cal A}}  \def\cC{{\cal C}} 
  \def\cF{{\cal F}} 
 \def\cH{{\cal H}}  
 \def\cK{{\cal K}} \def\cL{{\cal L}} 
\def\cM{{\cal M}} \def\cN{{\cal N}} \def\cO{{\cal O}} 
\def\cP{{\cal P}} \def\cQ{{\cal Q}} \def\cR{{\cal R}} 
 \def\cT{{\cal T}}

\def\R{{\mathbb R}} \def\C{{\mathbb C}} 

 \def\one{\mbox{1 \kern-.59em {\rm l}}}

\def\mso{\mathfrak{so}}

\newcommand{\Tr}{\mathrm{Tr}}
\newcommand{\tr}{\mathrm{tr}}
\newcommand{\End}{\mathrm{End}}
\newcommand{\Mat}{\mathrm{Mat}}

\def\hs{\mathfrak{hs}}

\def\Tr{\mbox{Tr}}

\def\NC{{\rm NC}}

\newcommand{\eq}[1]{(\ref{#1})}

 \allowdisplaybreaks[3]

\begin{document}



\renewcommand{\title}[1]{\vspace{10mm}\noindent{\Large{\bf#1}}\vspace{8mm}} 
\newcommand{\authors}[1]{\noindent{\large #1}\vspace{5mm}}
\newcommand{\address}[1]{{\itshape #1\vspace{2mm}}}


\begin{titlepage}
\begin{flushright}
 UWThPh-2023-9 \\
\end{flushright}
\begin{center}
\title{ {\Large One-loop effective action and emergent gravity  \\[1ex] on quantum spaces in the  IKKT matrix model}  }

\vskip 3mm

\authors{Harold C.\ Steinacker}

\vskip 3mm

\address{ 
{\it Faculty of Physics, University of Vienna\\
Boltzmanngasse 5, A-1090 Vienna, Austria  }  \\
Email: {\tt harold.steinacker@univie.ac.at}  
  }

\bigskip

\vskip 1.4cm

\textbf{Abstract}
\vskip 3mm

\begin{minipage}{14.5cm}%
\vskip 3mm

A detailed derivation of $3+1$  dimensional induced or emergent gravity in the IKKT matrix model at one loop is given, as announced in \cite{Steinacker:2021yxt}. 
The mechanism requires a brane configuration with structure $\cM^{3,1}\times \cK \subset \R^{9,1}$, where $\cM^{3,1}$ is the noncommutative space-time brane, 
and $\cK$ are compact fuzzy extra dimensions embedded in target space. 
The 3+1-dimensional Einstein-Hilbert action then arises in the one loop effective action
of the maximally supersymmetric IIB or IKKT matrix model, with
 effective Newton constant  determined by the Kaluza-Klein scale of $\cK$.
At weak coupling, all physical modes are confined to the brane, leading to $3+1$-dimensional low-energy physics.
The Einstein-Hilbert action can be interpreted as interaction of $\cK$ with the
space-time brane via IIB supergravity.
The vacuum energy does not act as cosmological constant, but stabilizes the 
brane structure at one loop.

\end{minipage}

\end{center}

\end{titlepage}

 \setcounter{tocdepth}{2}
\tableofcontents
%
%
\section{Introduction}

The present paper provides the detailed derivation and justification of the 
mechanism of emergent gravity at one loop presented in \cite{Steinacker:2021yxt}.

We consider the IKKT matrix model with background  given by some  matrix configuration
$\{X^a\}_{a=0,..,9}$, which can be interpreted as a quantum space or
brane $\cM\subset \R^{9,1}$ embedded in target space.
The one-loop contribution to the effective action is computed as a function of the background,
and written in terms of geometric structures in a suitable regime.
For suitable types of backgrounds with fuzzy extra dimensions,
this can then be related to the Einstein-Hilbert action amended with extra terms.

The main technical result is the explicit evaluation of the Gaussian integral over quadratic
fluctuations around the given background.
This boils down to the well-known trace-log formula for the gauge-fixed quadratic action.
In the IKKT model, the first three terms in the expansion of this formula
in the number of propagators cancel, due to maximal supersymmetry.
This is essential for obtaining an effective action which is finite on $3+1$-dimensional branes, without
any UV divergences. 
On the other hand, 
the leading non-trivial contribution is given by a formidable trace over
multiple commutators and various operators acting on the space of matrices.
The main challenge is  to evaluate this trace in a reasonable geometric way.
Even on simple homogeneous spaces, a traditional expansion in terms of
group-theoretical harmonics is unreasonable and intransparent. On non-trivial backgrounds without  symmetries, it may seem hopeless.

In the present paper, we show how this trace can be evaluated  using a novel technique 
in terms of
``string modes''. By definition, string modes are an over-complete set of rank-one matrices
of the form $|x\rangle\langle y|$,  where $|x\rangle$ are suitable (quasi-)coherent states on $\cM$.
These quasi-coherent states are optimally localized and coincide with the standard coherent states on quantized coadjoint orbits, but can be defined more generally.
String modes have been introduced first in \cite{Iso:2000ew}, and developed into a versatile tool in the context of
noncommutative field theory and matrix models in \cite{Steinacker:2016nsc,Steinacker:2022kji}.
The main tool for our purpose is a suitable trace formula, which is used
to evaluate the trace in a geometrical way.
The resulting one-loop effective action on $\cM$
can be written in terms of the
Einstein-Hilbert action, extended by extra ingredients including dilaton an
(gravitational) axion.
However, a more detailed physical assessment of the resulting gravity theory
is left to future work.

Apart from the technical computations which comprise the bulk of this paper, the most important
physical message is that (some) 3+1-dimensional gravity  arises on brane-like solutions of the matrix model
{\em at weak coupling} through quantum effects, without any compactification of target space.
In fact, all perturbative degrees of freedom propagate {\em on the brane}, and there is simply nothing that
could propagate in the bulk. This is not in conflict with string theory: $9+1$-dimensional IIB supergravity
can also be seen to arise in the bulk at one loop,
but only as interaction of branes in target space; if there is only one
brane as in the present setup, then this effect is only manifest as a short-range
$r^{-8}$ interaction of objects on the brane, which is sub-leading and negligible at
long distances. In other words: while IIB supergravity arises holographically in the bulk,
the gauge theory on the brane provides the appropriate description of physics at weak coupling,
leading to emergent gravity in terms of the effective (``open string'') metric on the brane.

On basic $3+1$- dimensional branes $\cM^{3,1} \subset \R^{9,1}$, the geometrical action induced at
one loop turns out to be some rather strange higher-derivative  action.
However on background branes of the structure  $\cM^{3,1} \times \cK_N \subset \R^{9,1}$
where $\cK_N$ is a quantized compact (``fuzzy'') space in transverse directions, the leading term of the
 induced gravitational action is indeed the Einstein-Hilbert term. This term can be seen as effective
IIB {\em interaction} of $\cM^{3,1}$ and $\cK_N$, and the gravitational coupling i.e. the Newton constant
is governed by the Kaluza-Klein (KK) scale defined by $\cK_N$.
It is essential that $\cK_N$  admits only finitely many degrees of freedom;
for a continuous space $\cK$, the one-loop computation would diverge, and the mechanism 
no longer works.
This may be the reason why such a mechanism has  not been seen in string theory up to now.
Such ``fuzzy extra dimensions'' have been studied extensively in the context of matrix models,
cf. \cite{Aschieri:2006uw,Steinacker:2014lma}.
They should not be confused with traditional string theory compactifications, which
may be implemented formally in  matrix models \cite{Connes:1997cr}, but would generically lead to UV divergences at one loop.

In a further step, these results are generalized
to backgrounds given by covariant quantum spaces $\cM^{3,1}$, which
 allow to mitigate the breaking of local Lorentz invariance by the
 symplectic structure. While the structure of the induced action is
 similar to the action for basic branes, the internal $S^2_n$ fibers over space-time
 leads to extra contributions to the one-loop effective action, which are
 computed explicitly.
 
 The last part of this work is focused on the
  1-loop contribution to the vacuum energy. This is interesting for several reasons:
  first, it provides the effective potential for the radius of the 
  compact space $\cK_N$, which is indeed seen to have a non-trivial minimum.
  This supports the viability of the present mechanism. Moreover, it turns out that
  the vacuum energy does not lead to a cosmological constant, but
  involves the symplectic volume form and the dilaton. 
  This is certainly very intriguing in view of the
  cosmological constant problem, however its significance remains to be understood.

\section{Heat kernel and induced gravity on commutative spaces}
\label{sec:heat-kernel}

It is well-known that quantum effects in field theory
covariantly coupled to a metric  lead to
gravity-type terms in the effective action, including notably the Einstein-Hilbert term,
as well as a cosmological constant. This ''induced gravity`` mechanism was pointed out first by Sakharov in 1967 \cite{Sakharov:1967pk}, and subsequently developed further from various points of view, cf.
\cite{Visser:2002ew}.
However, it typically leads to a huge cosmological constant set by the cutoff scale,
which constitutes the  notorious cosmological constant problem.
We will see that a variant of this mechanism arises within the IKKT model,
leading to a specific type of induced gravity. Remarkably, the cosmological constant problem
does not seem to arise in that framework.

\subsection{Trace formula on classical spaces}
\label{sec:traces-classical}


In the classical setting,
differential operators on configuration space can  be viewed as functions on phase space,
which suggests that their traces can be computed in terms of an integral
on phase space.  Choosing $\R^{2n}$ as configuration space,
such a trace formula can be obtained
using the following Gaussian wave packets
\begin{align}
 \psi^{(L)}_{k;y}(x) =  e^{i k x} e^{-(x-y)^2/L^2} \ ,
 \label{plane-wave-packets}
\end{align}
which are centered at $y$ with size $L$ and characteristic wave vector $k$.
Here $\R^{2n}$ is equipped with the standard Euclidean metric, and the
inner product of functions is defined as
\begin{align}
 \langle f,g\rangle = \int \frac{d^{2n} x}{(2\pi)^n} f^*(x) g(x) \ .
 \label{inner-packets-class}
\end{align}
Then their inner products are obtained as
\begin{align}
 \langle \psi^{(L)}_{k;x},\psi^{(L)}_{l;y} \rangle
 &= \frac{L^{2n}}{2^{2n}} \, e^{-\frac i2  (x+y) (k-l)} e^{ -\frac 12(x-y)^2/L^2 -\frac 18 L^2 (k-l)^2} \ .
 \label{inner-prod-wavepacket-class}
\end{align}
These $\psi^{(L)}_{k;y}$ form an over complete set of
functions in $\cH \cong L^2(\R^{2n})$ for any fixed $L$, and we will find analogous modes
$\Psi^{(L)}_{k;y}$ \eq{smeared-string} in the noncommutative case.

We would like to compute the trace $\Tr_\cH (\cO)$ for differential
operators $\cO = \cO(x,\del)$  acting on $\cH\cong L^2(\R^{2n})$.
This could be done using the plane wave basis
$e^{i k x}$ of $\cH$, however evaluating the trace for differential operators
with non-constant coefficients would be difficult and in-transparent.
A more useful and transparent approach is to use the above over-complete basis
$\psi^{(L)}_{k;y}$, choosing $L$ such that the coefficients of $\cO$
are approximately constant on the length scale $L$. We claim that
 the following general trace formula holds:
\begin{align}
\boxed{ \
 \Tr_\cH (\cO) = \frac{1}{(2\pi)^m}  
 \int d^{m} y \int d^{m} k \,
\frac{\langle\psi^{(L)}_{k;y}, \cO \psi^{(L)}_{k;y}\rangle }
{\langle\psi^{(L)}_{k;y}, \psi^{(L)}_{k;y}\rangle}\
 \ }
 \label{trace-wavelets}
\end{align}
for any $L$, where $m=\dim\cM=2n$. There are various ways to show this formula;
one possibility is to view the $\psi^{(L)}_{k;y}$  as position space wavefunctions
of  coherent states on the doubled Moyal-Weyl space $\R_\theta^{4n}$,
for $\theta = \frac 14L^2$. Then
\eq{trace-wavelets} reduces to the well-known completeness relation for coherent states,
which is basically a consequence of translation invariance
in position and momentum space $\R^{2n}_y \times \R^{2n}_k$:
Since $\cH$ is irreducible under this group,
Schur's Lemma implies that there is only one invariant functional on
$\End(\cH)$ up to normalization,
which is thus proportional to the above trace formula.
The normalization can be verified e.g. for
$\cO_0 = |\psi^{(L)}_{0;0}\rangle\langle\psi^{(L)}_{0;0}|$ using a simple Gaussian integral.

Note that the scale $L$ of the enveloping function in \eq{trace-wavelets} is
arbitrary. If $L$ is sufficiently large,  the $\psi^{(L)}$ are approximately plane waves,
which is useful to compute traces of differential operators with non-constant coefficients
as illustrated below.
On the other hand we can choose $L$ to be very short, to the point where
the $\psi^{(L)}$ are approximately delta functions. Then
the trace of integral operators $\phi(x) \to \int dy G(x,y)\phi(y)$
reduce to the integral $\int dx G(x,x)$
for the integral kernel.
Therefore the trace formula \eq{trace-wavelets} allows to interpolate between the position and
momentum point of view.

\subsection{Heat kernel and induced gravity}

The basic mechanism of induced gravity is best understood in terms of the heat kernel expansion\footnote{The
name ''heat kernel`` arises from the Euclidean case, where
$e^{-\a\Delta}$ is the fundamenal solution of the heat equation.},
and its generalization to the present framework. Consider
\begin{align}
H(\a):= \Tr e^{-\a\Delta} &= \sum_{n\geq 0} \a^{\frac{n-d}{2}} S_n
 \label{heat-kernel}
\end{align}
where $d$ is the dimension of $\cM$, and $\Delta$ is the
Laplacian on $\cM$ with metric $g_{\mu\nu}$;
more generally it could be any operator
which is bounded from below.
$H(\a)$ basically counts the number of eigenmodes of $\Delta$ below the
UV cutoff scale $\a^{-2} =: \L^2$.
If $\Delta$ was bounded (such as on compact fuzzy spaces), then $H(\a)$
would be analytic in $\a$.
On ordinary manifolds where $\Delta$ is unbounded from above,
$H(\a)$ can only be an asymptotic expansion, i.e. the above series
is formal and  diverges as $\a\to 0$.
The $S_n$ are local functionals known as Seeley-de Witt coefficients
\begin{align}
 S_n = \int d^dx \sqrt{|g|}\,\cL_n \ ,
\end{align}
which respect all (gauge and global) symmetries of $H(\a)$.
Typically, only terms with even $n$ are non-vanishing.
The leading terms  are as follows:
\begin{align}
 S_{0} &= \frac{1}{16\pi^2}\int d^4 x\sqrt{|g|}   \nn\\
 S_2 &= \frac{1}{16\pi^2}\int d^4 x\sqrt{|g|}\; \frac 16\cR[g]
 \label{induced-action-SdW}
\end{align}
etc.,
starting with the cosmological constant term for $n=0$,
and the Einstein-Hilbert action for $n=2$.
The $S_n$ are {\em local} 
functionals of the background metric and its derivatives up to order $n$, whose
explicit computation is similar to perturbative quantum field theory computations in the presence
of a UV cutoff $\L$ \cite{Vassilevich:2003xt,Gilkey:1995mj}.
In this sense, the heat kernel expansion provides the derivative expansion of the
generally non-local functional $\Tr\exp(-\a\Delta)$.
Once $H(\a)$ is known, the analogous expansion for expressions of the type
$\Tr\big(f(\Delta)e^{-\a\Delta}\big)$
can be obtained using a Laplace transformation.

As an application of the above trace formula \eq{trace-wavelets},
we compute the leading contribution to the heat kernel
on a topologically trivial 4-dimensional Riemannian manifold $(\cM,g)$
with Laplacian $\Delta$. Since the homogeneous  measure and inner product on $\R^4$
is assumed in this formula, we consider  $\cM$ as a homogeneous manifold $\R^4$,
equipped with a non-trivial metric $g_{\mu\nu}$.
We choose the length scale
\begin{align}
 L \gg  \frac{1}{k} \
 \label{k-L-regime}
\end{align}
such that
the relevant $\psi^{(L)}_{k;y}(x)$ are approximate local eigenfunctions of $\Delta_g$,
i.e.
\begin{align}
  \Delta_g\psi^{(L)}_{k;y}  \approx g^{\mu\nu}(x) k_\mu k_\nu \psi^{(L)}_{k;y}
  \label{Laplacian-planewave-class}
\end{align}
up to corrections suppressed by a factor $O(\frac{1}{k L})$ from the derivatives of the enveloping function in \eq{plane-wave-packets}.
We also assume that the metric  $g_{\mu\nu}$ is approximately constant over the length scale $L$;
this is easily
satisfied in the context of gravity.
Then
\begin{align}
  f(\Delta_g)\psi^{(L)}_{k;y}  \approx f(k \cdot k) \psi^{(L)}_{k;y}
\end{align}
where
\begin{align}
k\cdot k = k_\mu k_\nu g^{\mu\nu}(x)
\end{align}
may depend  on $x$. Similarly,
the $\psi^{(L)}_{k;y}$ are approximate local eigenfunctions of generic vector fields $v = v^\mu(x) \del_\mu$,
in the sense
\begin{align}
 v[\psi^{(L)}_{k;y}] =
v^\mu(x)\del_\mu\psi^{(L)}_{k;y} &\approx i k_\mu v^\mu  \psi^{(L)}_{k;y}  \nn\\
 v[v[\psi^{(L)}_{k;y}]] &\approx -(k_\mu k_{\nu}v^\mu v^\nu) \psi^{(L)}_{k;y}
 \label{VF-wavepackets}
\end{align}
etc.,
noting that $\del_x\psi^{(L)}_{k;y} \sim i k \psi^{(L)}_{k;y}$ for $|k| \geq \frac{1}{L}$ and assuming that $v$ is slowly varying.

We can now evaluate the heat kernel using a  Gaussian integration and
$\langle \psi^{(L)}_{k;x},\psi^{(L)}_{k;x} \rangle = \frac{L^{4}}{16}$  \eq{inner-prod-wavepacket-class}.
This gives
\begin{align}
\Tr e^{-\a\Delta}
&= \frac{1}{(2\pi)^4}  \int d^4 x\,\int d^4 k\, e^{-\a k\cdot k}
\ = \  \frac{1}{\a^2} \frac 1{(4\pi)^2}\,\int d^4 x\, \sqrt{g} \ .
\label{heatkernel-leading}
\end{align}
This is recognized as leading term $n=0$ in the heat kernel expansion \eq{heat-kernel},
or as  leading Seeley-de Witt coefficient \eq{induced-action-SdW}.
Traces over other functions of $\Delta$ can then be obtained
using a Laplace transformation in $\a$, including
UV-regulated traces in the context of QFT.
For example, the leading term of the 1-loop effective action induced by
integrating out a scalar field coupled to a background metric $g$ is obtained as
\begin{align}
 \Tr_\L \ln(\Delta)
&:= -\frac 12\Tr\int\limits_0^\infty\frac{d\a}{\a} e^{-\a\Delta-\frac 1{\a \L^2}}
=  -\frac 12\, \frac 1{(4\pi)^2}\,\int d^4 x\, \sqrt{g}
\int\limits_0^\infty\frac{d\a}{\a^3}e^{-\frac 1{\a \L^2}} \nn\\
&=  -\frac{ \L^4 }{2(4\pi)^2}\,\int d^4 x\, \sqrt{g} \,
\label{vacuum-free-cutoff}
\end{align}
up to sub-leading corrections in the  cutoff\footnote{The specific form of
the smooth UV-cutoff chosen here is convenient but not crucial.} $\L$.
The result is familiar from field theory, where it is interpreted as induced cosmological constant.
The computation  is however problematic, since the dominant
contribution  arises from the UV region near the cutoff $\L$;
this  will be avoided below due to supersymmetry.
The next, subleading term $S_2$ in the Seeley-de Witt expansion gives the
Einstein-Hilbert term \eq{induced-action-SdW},
with UV-divergent pre-factor $\L^2$ which sets the Newton constant.
We will see that in the IKKT model, the
Einstein-Hilbert term will arise as the leading contribution, and can be computed
directly using the formalism of string modes.

\section{String modes and a trace formula for quantum spaces}
\label{sec:string-modes}

Now consider a quantized compact symplectic space $\cM$ such as fuzzy $S^2_N$
\cite{hoppe1982QuaTheMasRelSurTwoBouStaPro,Madore:1991bw},
or a higher-dimensional analog such as a quantized coadjoint orbit, cf. \cite{Steinacker:2019fcb}.
The (noncommutative analog of) functions on such a space is given by
the algebra $\Mat(\cH)$ of operators on some Hilbert space $\cH$, which is
finite-dimensional for compact $\cM$.
The quantization map 
\begin{align}
 \cQ:\quad L^2(\cM) \to \Mat(\cH)
\end{align}
 then
allows to identify observables $\phi\in\Mat(\cH)$ with
 functions on $\cM$. In the example of $S^2_N$,
the noncommutative functions $\Mat(\cH)$ are spanned by the fuzzy
spherical harmonics $\hat Y^l_m = \cQ(Y^l_m)$, which are the quantizations of the
usual spherical harmonics. More generally, we will assume that the quantum space under consideration admits suitable (quasi-) coherent states $|x\rangle$ for $x\in\cM$, such that 
the completeness relation 
\begin{align}
 \int\limits_\cM\frac{\Omega}{(2\pi)^n} |x\rangle\langle x| = \one \
 \label{completeness-relation}
\end{align}
holds exactly (such as on quantized coadjoint orbits), cf. \cite{Steinacker:2020nva}.
Then a quantization map can be written down as
\begin{align}
 \cQ(\phi) = \int\limits_\cM\frac{\Omega}{(2\pi)^n} \, \phi(x) |x\rangle\langle x| \ .
\end{align}
We will consider a gauge theory on such a quantized space described
by some matrix model,  with background given by some matrix configuration $X^a$,
 $a=1,...,D$, interpreted as quantized embedding function of $\cM$ into target space:
 \begin{align}
  X^a \sim x^a: \quad \cM \hookrightarrow \R^D \ .
  \label{M-embedding-X}
 \end{align}
The matrix configuration defines a matrix Laplacian (or d'Alembertian)
\begin{align}
 \Box = [X^a,[X_a,.]]
\end{align}
acting on $\Mat(\cH)$, with indices contracted by the target space metric $\d_{ab}$ or $\eta_{ab}$;
for an introduction
see e.g. \cite{Steinacker:2010rh,Steinacker:2019fcb}.
Then for compact $\cM$, the analog of the heat kernel
\begin{align}
 H(\a) =  \Tr e^{-\a\Box}
 \label{heat-kernel-NC}
\end{align}
is analytic in $\a$.
The leading term in its series expansion is
\begin{align}
H(0) =  \Tr \one = \dim(\cH)
  =  \int\limits_\cM \frac{\Omega}{(2\pi)^n} \, =  {\rm Vol}_\Omega \cM   \ ,
\end{align}
which is finite and quantized for compact $\cM$, and has no dynamical significance whatsoever.
This  already
suggests that no cosmological constant problem should  arise in such a setting.
However, the significance of such expressions and
the relation with the commutative case is non-trivial due to UV/IR mixing.
This is best understood in terms of string modes,
which provide crucial  insights into the
UV regime of noncommutative field theory.

\paragraph{String modes.}

In the example of fuzzy $S^2_N$, the spherical harmonics $\hat Y^l_m$ are defined for
angular momentum $0 \leq l < N$. This quantum space admits $N$ area quanta,
with scale of noncommutativity set by $l_{\rm NC} = \sqrt{N}$.
Even though the  quantization map $\cQ$ is defined for all $l < N$,
 it is actually  misleading to think of modes with $l>l_{\rm NC}$
as quantized functions, because their multiplication 
is far from commutative. More generally, the modes  on generic quantum spaces $\cM$
with energy far above the semi-classical regime  
$L^{-1}_{\rm NC}$  should not be thought of as functions. This ''deep quantum`` regime is better described in terms of non-local
{\bf string modes} \cite{Iso:2000ew,Steinacker:2016nsc,Steinacker:2022kji},
which are bi-local modes in the algebra of
``functions" $\Mat(\cH)$  defined as
\begin{align}
\left|^x_{y} \right)
 &:=  |x\rangle\langle y|
 \qquad \in \Mat(\cH)  \
 \label{string-states}
\end{align}
where $|x\rangle, |y\rangle$ are (quasi-) coherent states on $\cM$.
\begin{figure}[h]
\begin{centering}
\includegraphics[width=0.35\textwidth]{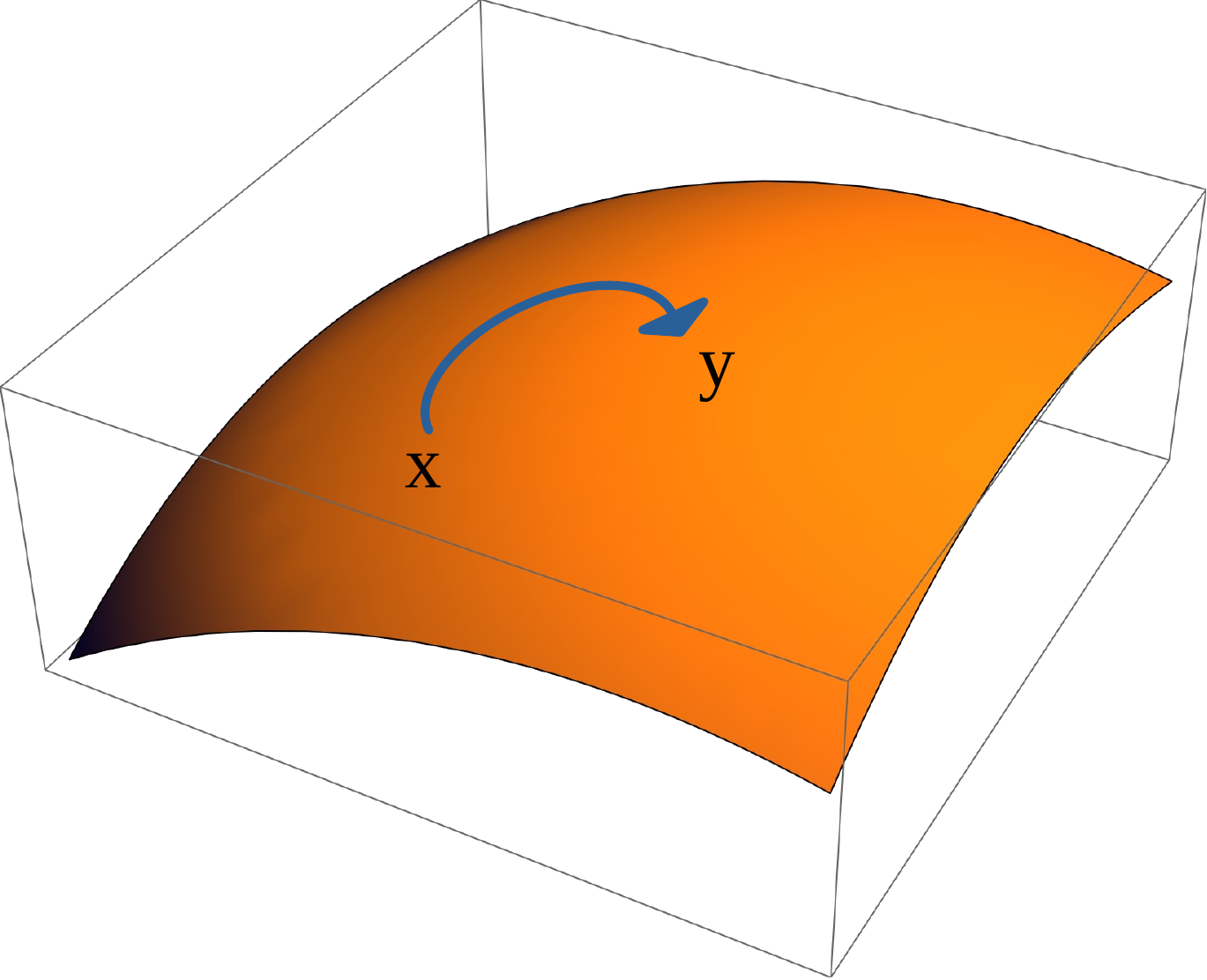}
\par\end{centering}
\caption{Visualization of a string mode $\left|^x_{y} \right)$ on a quantum space $\cM$. }
\label{fig:planar}
\end{figure}
We use the quantum mechanical ''ket`` notation to emphasize that
this is a normalized vector  in the Hilbert space $\Mat(\cH)$,
equipped with the (Hilbert-Schmidt) inner product
\begin{align}
 \langle F,G\rangle = \Tr(F^\dagger G) \ .
\end{align}
 Accordingly, the dual or adjoint string ''bra`` is denoted with $\left(^x_{y} \right|$.
It will be useful 
to identify the space of string modes  with the product space
\begin{align}
  \big\{\left|^x_{y} \right),\ \ x\in\cM, y\in \cM \big\} \ \cong  \cM_x \times \cM_y \ .
  \label{MM-st-identif}
\end{align}
This space inherits the symplectic (product) measure
from the symplectic measure  $\Omega$ on $\cM$.

The string modes have several remarkable properties.
One obvious property is bi-locality in $x$ and $y$, which is manifest in the inner product
\begin{align}
 \big(^x_{y} \big|^{x'}_{y'} \big) 
  &= \langle x|x'\rangle  \langle y'|y\rangle
  \ \approx \ e^{ - \frac 14 |x-x'|_g^2 -  \frac 14 |y-y'|_g^2} \ e^{i\varphi} \ .
  \label{string-states-inner}
\end{align}
 Here
\begin{align}
  |x-y|_g^2 = (x-y)^\mu (x-y)^\nu g_{\mu\nu} \ , \qquad g_{\mu\nu} = \frac{1}{L^2_{NC}}\d_{\mu\nu}
\end{align}
is the {\em quantum metric} on $\cM$ \cite{Steinacker:2020nva} in adapted coordinates, and
$\varphi$ is a gauge-dependent phase
\begin{align}
 \varphi(x,x';y,y') = -\frac 12 (x^\mu\theta^{-1}_{\mu\nu}{x'}^\nu - y^\mu\theta^{-1}_{\mu\nu}{y'}^\nu) \ ,
\end{align}
assuming the (approximate) localization properties of (quasi-) coherent states
on the Moyal-Weyl quantum plane\footnote{i.e.$[X^\mu,X^\nu] \approx i\theta^{\mu\nu}\one$, which can  be assumed to hold locally according to Darboux theorem.}
\begin{align}
 \langle x|y\rangle \approx
 e^{ - \frac 14 |x-y|^2_g} \ e^{-\frac i2 y^\mu\theta^{-1}_{\mu\nu}x^\nu} \ .
 \label{coherent-inner-explicit}
\end{align}
Most importantly, the string modes provide an over-complete basis of
$\Mat(\cH)$. Assuming that the (quasi-) coherent states on $\cM$ satisfy the
completeness relation \eq{completeness-relation}
and observing $\Mat(\cH) \cong \cH \otimes \cH^*$, we obtain
 the following (approximate or exact) completeness relation for string modes in $\Mat(\cH)$:
\begin{align}
 \one_{\Mat(\cH)}
  &= \int\limits_{\cM_x\times\cM_y}\! \frac{\Omega_x \Omega_y}{(2\pi)^{m}}\ \left|^x_{y} \right)\left(^x_{y} \right|\
 \label{complete-coh-End}
\end{align}
where $\dim\cM = m = 2n$. This formula is exact for any
quantized coadjoint orbit and the Moyal-Weyl quantum plane, as a
consequence of group invariance and Schur's Lemma. It is also applicable for generic deformations thereof,
because the underlying symplectic structure is  rigid, so that the undeformed symplectic form and coherent states can still be used.
As a consequence, we obtain the following trace relation 
 over hermitian matrices in $\Mat(\cH)$:
\begin{align}
 \Tr_{\Mat(\cH)} \cO &= \int\limits_{\cM\times\cM} 
  \frac{\Omega_x \Omega_y}{(2\pi)^m}
  \left(^x_{y} \right|  \cO \left|^x_{y} \right) \ .
  \label{trace-coherent-End-hermit}
\end{align}
This relation will be the basis for evaluating the one-loop effective action below.

\subsection{Kinematical properties}

In the context of matrix models, differential operators on quantum spaces are realized in terms of
commutators i.e. derivations. The matrix background $X^a$  defines
the following derivative
or ''matrix momentum`` operators acting on $\Mat(\cH)$:
\begin{align}
 \cP^a\, \phi &:= [X^a, \phi], \qquad \phi \in \Mat(\cH)   \nn\\
 \Box\, \phi  &:=   \cP^a \cP_a \phi \
 \label{P-P2-NC}
\end{align}
which  will play
an important role in the following.
In the semi-classical regime
i.e. for wavelength $> L_{\rm NC}$, the $\cP^a \sim i\{x^a,.\}$ are adequately interpreted as (Hamiltonian) vector fields on $\cM$.
For example on $\R^{2n}_\theta$, their eigenfunctions are given by plane waves $\phi= e^{i\tilde p X}$
\begin{align}
  \cP^a\,e^{i\tilde p X}  &= [X^a, e^{i\tilde p X}]
  = -\theta^{ab} \tilde p_b \, e^{i\tilde p X}
  = p^a\, e^{i\tilde p X} \
\end{align}
with $\tilde p_a =  p^b \theta^{-1}_{ba}$.
However  outside of the semi-classical regime,
this  interpretation is inappropriate, and operators such as
$\cP^a$ are in general {\em nonlocal}.
Although this non-locality is only significant  in the UV regime,
it plays a crucial role in quantum field theory on quantum spaces, due to virtual  
UV modes propagating in the loops.
This becomes manifest in terms of the string modes. For example,
the  matrix elements
\begin{align}
\left(^x_{y} \right|  \cP^a \left|^{x'}_{y'} \right)
 &\approx ({\bf x^a}(x) -  {\bf x^a}(y))
 \langle x | x'\rangle \langle y' | y\rangle  \nn\\
  \left(^x_{y} \right|  \cP^a \left|^x_{y} \right)
 &= {\bf x^a}(x) -  {\bf x^a}(y) \ \approx x^a - y^a \
 \label{Pa-exp}
 \end{align}
  of $\cP^a$
are approximately diagonal, using $\langle x|X^a|x'\rangle \approx {\bf x^a}(x)\langle x|x'\rangle$ where
\begin{align}
{\bf x^a}(x):= \langle x|X^a|x\rangle \ \approx \ x^a \ .
    \label{M-embedding-symbol}
\end{align}
The Laplacian is also approximately diagonal, with
\begin{align}
\left(^x_{y} \right| \Box \big|^{x'}_{y'} \big)
 = \left(^x_{y} \right| \cP^a \cP_a \big|^{x'}_{y'} \big)
  &\approx  E_{xy} \, \langle x | x'\rangle \langle y' | y\rangle \ , \nn\\
\left(^x_{y} \right| \cP^a\cP_a \left|^x_{y} \right)
 &=  E_{xy} \ .
  \label{propagator-general}
\end{align}
Here
 \begin{align}
 E_{xy} &= (\vec{\bf x}(x) -\vec{\bf x}(y))^2 +  \Delta^2_x + \Delta^2_y
 \label{PP-strings}
\end{align}
can be interpreted as energy of a string mode, which is  given by its length square
plus the intrinsic uncertainty
$\Delta^2_{x} = \sum_a\langle x|(X^a - {\bf x^a}(x))^2|x\rangle \approx L_{\rm NC}^2$
of the  coherent states.
 These results can be subsumed as follows
\begin{align}
 \cP^a \big|^{x}_{y} \big) \approx (x-y)^a\big|^{x}_{y} \big), \qquad
 \Box\big|^{x}_{y} \big) \approx E_{xy}\big|^{x}_{y} \big) \ .
 \label{propagator-approx}
\end{align}
Hence
the string mode $\left|^{x+y/2}_{x-y/2} \right)$ 
is an approximate eigenstate of $\cP\approx y$ localized near $x$.
This suggests that it should be interpreted as plane wave packet.
To make this precise,
we should  distinguish between a semi-classical and a ''deep quantum regime`` for string modes:

\subsection{Short string modes as localized wave packets}

We will denote string modes $\big|^{x}_{y} \big) $
with $|x-y| \leq L_\NC$ as {\bf short string modes}. They provide the NC analog of optimally localized
wave packets around $x\approx y$, with characteristic size $L_\NC$
and linear momentum  determined by $x-y$.
They also  mark the boundary between the classical IR and
stringy UV regime in $\Mat(\cH)$.
To see this, consider a 4-dimensional quantum space
$\cM$, which we assume to satisfy the
  almost-K\"ahler condition
  \begin{align}
 \theta^{-1}_{\r \mu} \theta^{-1}_{\s \nu} g^{\r \s} = g_{\mu\nu} \
 \label{almost-Kahler}
\end{align}
which holds on $\R^{4}_\theta$.
We can then identify it locally with a quantum plane as long as
$\theta^{\mu\nu} \approx const$,
\begin{align}
 \cM \cong \R^4_\theta \ ,
 \label{TM-Rn-identif}
\end{align}
 such that their quantum metrics $g$ coincide.
 Using the Wigner map\footnote{Recall that the Wigner map is the inverse of the Weyl quantization map, which is
an isometry from $L^2(\R^{2n})$ with measure $\frac{\Omega}{(2\pi)^n}$ to  Hilbert-Schmidt operators, defined by $e^{i k x} \mapsto e^{i k X}$.}, we obtain the  following
 isometric identification of short string modes on $\cM \approx \R^4_\theta$ with Gaussian wave-packets on $\R^4$:
 \begin{align}
  \boxed{\
 e^{\frac i2 k^\mu \theta^{-1}_{\mu\nu} y^\nu}\big|^{y+\frac k2}_{y-\frac k2} \big) \
 =: \Psi_{\tilde k;y} \ \cong \ \psi_{\tilde k;y}(x) := 4\, e^{i \tilde k x} e^{-|x-y|_g^2} \
\ }
\label{string-wavepacket-identif}
 \end{align}
 recalling $\tilde k_\mu = k^\nu \theta^{-1}_{\nu\mu}$.
 To verify this, we compute  the  inner product for the Gaussian modes as
\begin{align}
 \langle \psi_{\tilde k;x},\psi_{\tilde l;y} \rangle
 &= \int \frac{d^4z}{(2\pi L_\NC^2)^2}\, \psi^*_{\tilde k;x}(z)\psi_{\tilde l;y}(z)
  =  e^{\frac i2(\tilde l - \tilde k)(x+y)}
    e^{-\frac 12|x-y|^2_g - \frac 1{8}|k-l|^2_g }
 \label{inner-prod-wavepacket}
\end{align}
 using the almost-K\"ahler relation $|\tilde k|_g^2 = |k|_g^2$ \eq{almost-Kahler}. This
agrees with  the inner product \eq{string-states-inner} of the short string modes
\begin{align}
  \langle\Psi_{\tilde k;x} ,\Psi_{\tilde l;y}\rangle
   &= \ e^{ - \frac 14 |x+\frac 12 k-y-\frac 12 l|_g^2
   -  \frac 14 |x-\frac 12 k-y+\frac 12 l|_g^2} \
   e^{\frac i2 l^\mu \theta^{-1}_{\mu\nu} y^\nu} e^{-\frac i2 k^\mu \theta^{-1}_{\mu\nu} x^\nu}
   e^{-\frac i2\big((x+\frac k2)^\mu(y+\frac l2)^\nu + (y-\frac l2)^\mu(x-\frac k2)^\nu \big)\theta^{-1}_{\mu\nu}} \nn\\
   &=   e^{\frac i2 (\tilde l - \tilde k)(x + y)} \,
   e^{-\frac 12|x-y|_g^2 - \frac 18 |k-l|_g^2} \ ,
 \label{inner-prod-string-short}
\end{align}
confirming the identification \eq{string-wavepacket-identif}.
It may be illuminating to make the link with Quantum Mechanics, where coherent states can be identified with Gaussian wave-functions in position space. In contrast,
the string modes $\left|^x_{y} \right)$  are operators,
which correspond to Gaussian functions on phase space.

Refining the  identification \eq{MM-st-identif},
we can now identify the space of short string modes  with the cotangent bundle
\begin{align}
 (\cM \times \cM)_{\rm short} &\cong T^*\cM  \nn\\
  (x,y) &\mapsto (x,\tilde k),
  \qquad \tilde k_\nu = (y-x)^\mu \theta^{-1}_{\mu\nu} \ ,
  \label{MM-TM-identif}
\end{align}
assuming that  $\cM$ is approximately flat on scales $|y-x|<L_{\NC}$.
The symplectic (product) measure on $\cM\times\cM$ then reduces to the canonical
measure on $T^*\cM$
\begin{align}
\Omega_x \Omega_y = \sqrt{\theta^{-1}_{\mu\nu}}\,d^{m} x
 \sqrt{\theta^{-1}_{\mu\nu}}\, d^{m} y
 = d^{m} x\, d^{m}\, \tilde k \ .
 \label{measure-TsM-shortss}
\end{align}
The  quantum metric is an extra structure on $\cM$ and $T_x\cM$,
which is encoded in the inner product
\eq{string-states-inner}
of the short string modes.

\subsection{Semi-classical Gaussian modes}
\label{sec:semi-class-Gauss}

We have seen that  short string modes can be viewed as
localized Gaussian wave packets,
which are just outside of the semi-classical regime.
This suggests that semi-classical wavefunctions can be obtained as
Gaussian averages of short string modes over larger scales $L \gg  L_\NC$:
 \begin{align}
  \Psi^{(L)}_{k;y} &:=
   \int  \frac{d^4 z}{(2\pi L_\NC^2)^2}\, e^{-|y-z|^2/L^2}\Psi_{k;z}
  \label{smeared-string}
 \end{align}
  where
 \begin{align}
  |y-z|^2  \equiv |x-y|^2_g\, L_\NC^2 \ = (x-y)^\mu (x-y)^\nu \d_{\mu\nu} \ .
 \end{align}
  They provide a useful (over-complete) basis of semi-classical modes,
  assuming  $\cM \sim\R^4_\theta$ over the scale $L$.
According to \eq{string-wavepacket-identif}, $\Psi^{(L)}_{k;y}$
 is  identified isometrically with the wavefunction
 \begin{align}
  \psi^{(L)}_{k;y}(x) &=
   \int \frac{d^4 z}{(2\pi L_\NC^2)^2}\, e^{-|y-z|^2/{L}^2}  \psi_{k;z}(x)
   \  \stackrel{L\gg L_\NC}{\approx} \   e^{i k x} e^{-|x-y|^2/L^2} \ .
 \label{smeared-string-wavefunct}
 \end{align}
For $L\gg L_\NC$, this agrees indeed with the symbol of these semi-classical modes
\begin{align}
 \langle x|\Psi^{(L)}_{k;y}|x\rangle
 &= \int \frac{d^4 z}{(2\pi L_\NC^2)^2}\,  e^{-|y-z|^2/L^2}
 \langle x|\Psi_{k;y}|x\rangle
  \  \stackrel{L\gg L_\NC}{\approx} \  e^{i k x} e^{-|x-y|^2/L^2} \
\end{align}
 assuming that $\cM$ is sufficiently flat near $y$ and
 $k^2 \ll \frac 1{L_{NC}^2}$.
We conclude that the $\Psi^{(L)}_{k;y}$ can be interpreted
for small $k$ as
semi-classical Gaussian
modes  with  size $L \gg L_\NC$, 
which approximate plane waves
 $e^{i k X} \sim e^{i k x}$.
Their momentum or wavenumber is measured by
\begin{align}
 \cP^a \Psi^{(L)}_{\tilde k;y} \approx k^a  \Psi^{(L)}_{\tilde k;y} ,
 \qquad   i\{x^a,\psi^{(L)}_{\tilde k;y}\} \approx k^a \psi^{(L)}_{\tilde k;y} \ .
 \label{momentum-WP}
\end{align}
 Since  \eq{string-wavepacket-identif} is an isometry,
  the inner products of $\Psi^{(L)}_{k;y}$
and $\psi^{(L)}_{k;y}$ coincide, and
are given  for large $L$ by
\begin{align}
  \langle\Psi^{(L)}_{k;x} ,\Psi^{(L)}_{l;y}\rangle
  &= \int \frac{d^4z}{(2\pi L_\NC^2)^2} \psi^{(L)}_{k;x}(z) \psi^{(L)}_{l;y}(z) \nn\\
  &\stackrel{L\gg L_\NC}{\approx} \  \frac {L^4}{16 L_\NC^4}\, e^{\frac i2 (l - k)_\mu(x + y)^\mu} \,
   e^{-\frac 12|x-y|^2/L^2 - \frac 18 |k-l|^2 L^2} \ .
 \label{inner-prod-string-long}
\end{align}
For large $L$, this approaches the $\delta(k-l)$ orthogonality relation for plane waves.

%

\paragraph{UV-convergent traces on quantum spaces.}

Now assume that $\Tr \cO$ is UV-finite, which 
is the case e.g. for $\cO = (\Box + m^2)^{-n}$ for sufficiently large $n$.
Using the correspondence \eq{string-wavepacket-identif},
this implies that the dominant part of the $y$ integral \eq{trace-coherent-End-hermit} originates from
some  ball $B_L(x)$ with radius $L$ around $x$, which we assume to be sufficiently small
so that the identification \eq{measure-TsM-shortss} and \eq{MM-TM-identif}  with $T^*\cM$
 can be used. Then
\begin{align}
  \Tr \cO 
 &\approx \frac{1}{(2\pi)^{m}}
 \int\limits_{\cM} \Omega_x \int\limits_{B_L(x)} \Omega_y
  \left(^{y}_{x} \right| \cO \left|^{y}_{x} \right) \nn\\
  &\approx \frac{1}{(2\pi)^{m}} \int\limits_{T^*\cM} d^m x d^m k
  \langle \Psi_{k,x}, \cO  \Psi_{k,x}\rangle 
   = \frac{1}{(2\pi)^{m}} \int\limits_{\cM} d^m x\sqrt{G}\!\int\!\!\frac{d^m k}{\sqrt{G}}
  \langle \Psi_{k,x}, \cO  \Psi_{k,x}\rangle  \ .
 \label{trace-coherent-End-short}
\end{align}
 (dropping the tilde on $k$), which is made more
transparent in terms of an arbitrary metric $G_{\mu\nu}$ on $\cM$.
No cutoff in $k$ is needed due  to the assumed UV finiteness.
We can then effectively work on the local $\cM \approx \R^{2n}_\theta$,
and replace the short string modes $\Psi_{k,x}$
with the semi-classical modes $\Psi^{(L)}_{k;y}$  \eq{smeared-string-wavefunct}
with characteristic size $L \gg 1/k \gg L_\NC$.
Then the above trace formula
generalizes immediately to the semi-classical string modes
\begin{align}
  \Tr \cO 
 &\approx \frac{1}{(2\pi)^m}\!\int\limits_{\cM} d^m x \sqrt{G}
 \int\! \frac{d^m k}{\sqrt{G}}
  \frac{\langle \Psi^{(L)}_{k,x}, \cO  \Psi^{(L)}_{k,x}\rangle}
  {\langle \Psi^{(L)}_{k,x}, \Psi^{(L)}_{k,x}\rangle}
  \label{trace-NC-conv-int}
\end{align} 
for any $L$. We have thus recovered precisely
the classical trace formula \eq{trace-wavelets}
 within the present quantum framework.
 All these formulas 
 can be understood as a consequence of the
isometry arising from  the equivalent Gaussian inner product structures 
\eq{inner-prod-string-short} and \eq{inner-prod-string-long}.
They are equivalent to a completeness
relation, which for homogeneous spaces \footnote{If $\cM$ is not homogeneous,
the argument generalizes using a partition of unity argument, for suitable scales.} 
follows directly from translation invariance.

\paragraph{Laplacian and effective metric.}

 Generalizing \eq{momentum-WP}
 to the matrix d'Alembertian $\Box= \eta_{ab}\cP^a \cP^b$ for target space with Minkowski signature, we obtain the analog of \eq{Laplacian-planewave-class}
 \begin{align}
  \Box \Psi^{(L)}_{\tilde k;y}  \approx
   |k|^2_\eta\Psi^{(L)}_{\tilde k;y}
   \approx \cQ(|\tilde k|^2_\g \psi^{(L)}_{\tilde k;y})
   \approx \cQ(-\g^{ab}\del_a \del_b\psi^{(L)}_{\tilde k;y} )  \
    \approx\cQ(\Box_\g \psi^{(L)}_{\tilde k;y})
  \label{Box-gamma-psi}
 \end{align}
 provided $\frac 1{L_\NC} \gg k \gg 1/L$.
 Here $\cQ$ is the (Weyl) quantization map, and
 \begin{align}
 \g^{ab} &:= \theta^{aa'}\theta^{bb'}\eta_{a'b'} \ ,  \nn\\
  |k|^2_\eta &=  \eta_{ab}k^a k^b =  \g^{ab} \tilde k_a\tilde k_b
  = |\tilde k|^2_\g \ ,
 \label{eff-gamma-1}
\end{align}
 assuming that $\del\theta^{ab}$ is negligible.
 Hence in that regime,  $\g^{ab}$ is the effective metric
 underlying the Matrix d'Alembertian $\Box$.
 The effective metric will be recognized as a conformal
rescaling $G^{ab} = \r^{-2} \g^{ab}$ of $\g^{ab}$.
 In string theory, this
 is known as open string metric on a brane with $B$ field, which governs the
 gauge theory on the brane.

\subsection{Long string modes}

Finally,  the {\bf long string modes}  $\big|^{x}_{y} \big) $
for $|x-y| > L_\NC$
are completely non-local objects on $\cM$,
which have no analog on classical spaces.
 They are naturally viewed as dipoles or strings linking $x$ with $y$
which are bi-local in position space, and
 have good localization properties in momentum space as well.
This makes them   novel and  very interesting objects from a QFT point of view.
They provide the appropriate description of the UV or deep quantum
regime of NC field theory, and are responsible for UV/IR mixing.
Note that the long string modes comprise the bulk of $\Mat(\cH)$:
e.g. on the fuzzy sphere $S^2_N$, the dividing line between the
semi-classical regime and the deep quantum regime is given by
the angular momentum $l\sim\sqrt{N}$, which is far below the UV cutoff at
$l_{UV} = N-1$. The short string modes define the boundary between these
regimes.

\paragraph{Noncommutative heat kernel revisited.}

We have seen
that the UV sector of  noncommutative spaces is governed by string modes, which are
highly non-local. These will dominate in any (would-be) UV-divergent trace over modes, and
invalidate the classical results. Hence the heat kernel \eq{heat-kernel-NC}
will approximate the
commutative one only for large $\a > \L^{-2}_{\NC}$, where the semi-classical modes contribute;
then the matrix d'Alembertian approximates the metric one  as
 $\Box \sim \r^2\Box_G$ \cite{Steinacker:2010rh}.
In contrast for small $\a \to 0$, the non-local UV regime
will dominate and lead to a totally non-local  heat kernel.
Hence the classical behavior of the heat kernel should be recovered
for the IR regime $\a > \L^{-2}_{\NC}$,
rather than  the UV regime $\a\to 0$ where non-local effects will dominate.
This can indeed be verified in  non-commutative field theory \cite{Blaschke:2010rr}.

We conclude  that a reasonably gravity theory on noncommutative spaces
can only be expected from a UV-finite theory, notably from the IKKT model on a $3+1$-dimensional brane.
Rather than focusing on the heat kernel $H(\a)$, we will then
compute  the one loop effective action, and extract the gravitational action directly.
This will  indeed lead  to an induced Einstein-Hilbert term similar to
$S_2$  \eq{induced-action-SdW}, but only in the
presence of fuzzy extra dimensions, since otherwise it would be cancelled by SUSY.

\section{The IKKT model and quantum space-time}

We can now study the effective gravity on $3+1$-dimensional noncommutative or quantized branes as described by the
IKKT or IIB matrix model \cite{Ishibashi:1996xs}. This model
is defined by the action
\begin{align}
S[T,\Psi] = \frac{1}{g^2}{\rm Tr}\big( [T^a,T^b][T_{a},T_{b}] 
\,\, + \overline\Psi \Gamma_a[T^a,\Psi] \big) \
\label{MM-action-IKKT}
\end{align}
where $T^a \in \Mat(\cH), \ a=0,...,9$ are hermitian matrices, and
 $\Psi$ are Majorana-Weyl spinors of $SO(9,1)$
  whose entries are (Grassmann-valued) matrices.
 Indices are contracted with the $SO(9,1)$ -invariant tensor $\eta^{ab}$, which is interpreted as metric on target space $\R^{9,1}$.
This model admits a manifest $SO(9,1)$ symmetry acting as
\begin{align}
 T^a \to \L(g)^a_b T^b\,,\quad  &\Psi_\a \to \tilde \pi(g)_\a^\b \Psi_\b\, ,
\end{align}
where $\tilde \pi(g)$ denotes the spinorial representation
of the universal covering group $\widetilde{SO}(9,1)$, and $\L(g)$ denotes the vector representation.
As a matrix model, the action is also invariant under
\begin{align}
 T^a \to U^{-1} T^a U\,,\qquad  &\Psi \to U^{-1} \Psi U\,
\end{align}
for $U \in U(\cH)$.
This is treated as a gauge symmetry, which means that configurations related by
such transformations are identified.
Moreover the model enjoys maximal supersymmetry,
as the action is
nothing but the dimensional reduction of $U(\cH)$ $\cN=1$ Super-Yang-Mills
on $\R^{9,1}$ to a point. The explicit SUSY transformations will not be needed, but it is responsible for the good behavior under quantization, which is defined by the ''path`` integral
over the space of matrices.

\subsection{Background and geometric structures}

Consider an ''almost-commutative`` background\footnote{It can be shown quite generally that  almost-commutative backgrounds -- i.e. with small commutators $[T^a,T^b]$ -- admit an interpretation as
a quantized symplectic brane \cite{Steinacker:2020nva}. }
defined by some matrix configuration $T^a$, interpreted as
quantized embedding map of some brane $\cM$:
 \begin{align}
  T^a: \quad \cM \hookrightarrow \R^D \ .
  \label{M-embedding-T}
 \end{align}
 We will routinely identify matrices with functions in the semi-classical regime via $\cQ$.
 The matrices $T^{a}$ define a frame as follows
\begin{align}
E^{\dot a} &:= -i[T^{\dot a},.] \sim \{ T^{\dot a},.\}   \nn\\
 E^{\dot a \mu} &= \{T^{\dot a},x^\mu\}  \label{frame-general}
\end{align}
where $x^\mu$ are local coordinates on $\cM$.
Sometimes a dot is used to distinguish frame indices from coordinate labels.
Observe that the frame is always given by some Hamiltonian vector field on the brane $\cM$
in the present framework. It governs the propagation of all fluctuations in the
 matrix model, as the kinetic terms have the structure
\begin{align}
  \Tr([T^{\dot a},\Phi][T_{\dot a},\Phi]) \sim \int d^4x\sqrt{|G|}\, G^{\mu\nu}\del_\mu\phi\del_\nu\phi \
\end{align}
with the {\bf effective metric} on $\cM$
\begin{align}
G^{\mu\nu} &= \r^{-2}\, \g^{\mu\nu} , \qquad
 \g^{\mu\nu} = \eta_{{{\dot a}}{\dot b}} \tensor{E}{^{\dot a}^\mu} \tensor{E}{^{\dot b}^\nu} \
  \label{eff-metric-def}
\end{align}
for a uniquely determined dilaton $\rho$. Using partial integration, one can read off the 
following relation with the metric d'Alembertian \cite{Steinacker:2010rh}
\begin{align}
 \Box = [T^a,[T_a,.]] \sim  -\{T^a,\{T_a,.\}\} = \r^2 \Box_G \ .
 \label{Box-metric}
\end{align}
Given this frame,
it is natural to consider the associated Weizenb\"ock
connection $\nabla^{(W)} E^{\dot a} = 0$, which is flat but has non-trivial torsion
given by \cite{Steinacker:2020xph}
\begin{align}
 \tensor{T}{^{\dot a}^{\dot b}^\mu} &:= \{\cF^{\dot a\dot b},x^\mu\}
  = -\tensor{E}{^{\dot a}^\nu}\del_\nu\tensor{E}{^{\dot b}^\mu}
  + \tensor{E}{^{\dot b}^\nu}\del_\nu\tensor{E}{^{\dot a}^\mu}
 \label{torsion-general}
\end{align}
using  the Jacobi identity. Covariantizing the frame indices using 
$\tensor{E}{^{\dot a}^\mu}$ and rising resp. lowering the indices  with 
$\g^{\mu\nu}$ then gives
\begin{align}
  \tensor{T}{_\s_\k^{\dot a}}
  &= \del_\s\tensor{E}{^{\dot a}_\k} - \del_\k\tensor{E}{^{\dot a}_\s} \ ,
   \label{T-dE}
\end{align}
i.e. the torsion is the exterior derivative of the coframe.
Its totally antisymmetric component  $T^{(AS)\nu\rho \mu} = T^{\nu\rho \mu} + cycl.$ then
defines an ''axionic`` vector field $\tilde T_{\sigma}$ via
\begin{align}
 \tensor{T}{^{(AS)}^\nu_\r_\mu} &=: -\sqrt{|G|}G^{\n\n'}\varepsilon_{\n'\r\mu\s} G^{\s\s'}
 \tilde T_{\s'} \ .
 \label{T-AS-dual-2}
\end{align}
One can show using the eom of the matrix model that
the corresponding one-form is exact \cite{Fredenhagen:2021bnw} and can be written as
\begin{align}
 \r^2 \tilde T_\mu = \del_\mu \tilde\r \ 
 \label{T-del-onshell}
\end{align}
for some function $\tilde\r$.
This function will be denoted as {\em axion}.

Finally, we note that
the Ricci scalar for the metric $G_{\mu\nu}$ can be expressed in terms of the 
above Weitzenb\"ock torsion using the identity (E.2) in \cite{Fredenhagen:2021bnw} 
 \begin{align}
   \cR &=  - \frac 12\tensor{T}{^\mu_\s_\r} \tensor{T}{_\mu_{\s'}^\r} G^{\s\s'}
  - \frac 12\tilde T_{\nu} \tilde T_{\mu}  G^{\mu\nu}  
        + 2 \r^{-2} G^{\mu\nu}\del_\mu\r\del_\nu\r 
        - 2\nabla_{(G)}^\mu (\r^{-1} \del_\mu\r) \ .
        \label{R-offshell-T-2}
 \end{align}
 This type of identity is well-known in the context of teleparallel gravity \cite{Aldrovandi:2013wha},
 but the detailed relation with the dilaton and axionic vector field is particular to the present framework.

\subsection{Evaluation of the 1-loop action using string modes}

We will restrict ourselves to the one-loop effective action, which
is defined by a Gaussian integral around some given matrix background $\bar T^a$,
taking into account the fermions and the ghosts due to gauge-fixing:
\begin{align}
 \int\limits_{\rm 1\, loop} dT d\Psi d\bar c dc\,
 e^{iS[T,\Psi,c]}
 = e^{i (S_0[\bar T] + \Gamma_{\!\rm{1 loop}}[\bar T]) }
  = e^{i \Gamma_{\rm eff}[\bar T]} \ .
 \end{align}
The oscillatory integral over matrices is defined by adding the following term to the
action 
\begin{align}
 S \to  S + i \varepsilon {\rm Tr}\sum_a (T^a)^2
\end{align}
which amounts to Feynman's $i\varepsilon$ regularization, 
cf. \cite{Karczmarek:2022ejn,Hirasawa:2022qzg}.
Carrying out the Gaussian integrals,
one obtains the following formula \cite{Ishibashi:1996xs,Chepelev:1997av,Blaschke:2011qu}
\begin{align}
\Gamma_{\!\textrm{1loop}}[T]\!
&= \frac i2 \Tr \Big(\log(\Box-i\varepsilon  - \Sigma^{(\cA)}_{ab}[ \cF^{ab},.])
-\frac 12 \log(\Box-i\varepsilon - \Sigma^{(\psi)}_{ab}[ \cF^{ab},.] )
- 2 \log (\Box-i\varepsilon )\Big)   \nn\\
 &= \frac i2 \Tr \Bigg(\sum_{n>0} \frac{1}n \Big(-\big((\Box -i\varepsilon )^{-1}\Sigma^{(\cA)}_{ab}[ \cF^{ab},.]
   \big)^n
  \, +\frac 12 \big((\Box -i\varepsilon )^{-1}\Sigma^{(\psi)}_{ab}[ \cF^{ab},.]\big)^n \Big)  \Bigg) \nn\\
  &= \frac i2 \Tr \Big(-\frac 14 \big((\Box -i\varepsilon )^{-1}\Sigma^{(\cA)}_{ab} [ \cF^{ab},.] \big)^4
  +\frac 18 \big((\Box -i\varepsilon )^{-1}\Sigma^{(\psi)}_{ab} [ \cF^{ab},.]\big)^4 \,\, +  \cO(\Box^{-1}[ \cF^{ab},.])^5 \! \Big)
\label{Gamma-IKKT}
\end{align}
dropping the terms with $n\geq 5$.
Here
\begin{align}
 \cF^{ab} &= i[T^a,T^b] \ ,   \nn\\
  (\Sigma_{ab}^{(\psi)})^\a_\b &= \frac i{4} [\Gamma_a,\Gamma_b]^\a_\b \ ,  \nn \\
     (\Sigma_{ab}^{(\cA)})^c_d &= i(\d^c_b \d_{ad} - \d^c_a \d_{bd}) \ ,
\label{Sigma-A-spinor}
\end{align}
where $(\Sigma_{ab})^\a_\b$ are $SO(9,1)$ generators acting on the spinor or vector representation, respectively.
The  leading terms in this expansion cancel  due to maximal supersymmetry,
so that  the first non-vanishing term arises for $n=4$, given by
\begin{align}
 \Gamma_{\!\textrm{1loop};4}[T]\! &=
 \frac i8 \Tr \Bigg(- ((\Box -i\varepsilon )^{-1}(\Sigma^{(\cA)}_{ab} [ \cF^{ab},.])^4
  +\frac 12 ((\Box -i\varepsilon )^{-1}\Sigma^{(\psi)}_{ab} [ \cF^{ab},.])^4 \Bigg) \nn\\
  &= \frac i4 \Tr\Big((\Box -i\varepsilon )^{-1}[\cF^{a_1 b_1}, \ldots (\Box -i\varepsilon )^{-1}[\cF^{a_4 b_4},.]]]]\Big) \nn\\
 &\quad  \big(-4 \eta_{b_1 a_2} \eta_{b_2 a_3} \eta_{b_3 a_4} \eta_{b_4 a_1}
- 4 \eta_{b_1 a_2} \eta_{b_2 a_4} \eta_{b_4 a_3} \eta_{b_3 a_1}
- 4 \eta_{b_1 a_3} \eta_{b_3 a_2} \eta_{b_2 a_4} \eta_{b_4 a_1} \nn\\
&\quad +  \eta_{b_1 a_2} \eta_{b_2 a_1} \eta_{b_3 a_4} \eta_{b_4 a_3}
 +  \eta_{b_1 a_3} \eta_{b_3 a_1} \eta_{b_2 a_4} \eta_{b_4 a_2}
+  \eta_{b_1 a_4} \eta_{b_4 a_1} \eta_{b_2 a_3} \eta_{b_3 a_2} \big) \ .
\label{Gamma-IKKT-4}
\end{align}
This vanishes  for
constant fluxes $\cF^{ab}$, and more generally for 
identical parallel $\R^{2n}_\theta$ branes
due to maximal SUSY.
For generic branes with different fluxes,  a
residual interaction arises, consistent with  IIB supergravity.
We will compute $\Gamma_{\!\textrm{1loop};4}[T]$  in the weakly coupled regime,
 on a background describing  a $3+1$-dimensional brane $\cM^{3,1}$
interpreted as space-time,
and on  $\cM^{3,1} \times \cK \ \subset \R^{9,1}$
describing  a space-time brane  with compact fuzzy extra dimensions $\cK$.

The string modes are very useful to evaluate \eq{Gamma-IKKT-4}, because they
have good localization properties in {\em both} position and momentum
as discussed above.
In particular,
\begin{align}
 (\Box -i\varepsilon)^{-1}\left|^x_{y} \right)
 &\sim \frac 1{|x- y|^2 + 2\Delta^2 -i\varepsilon} \ \left|^x_{y} \right) \nn\\
 (\Box -i\varepsilon)^{-1}[ \cF^{ab},.] \left|^x_{y} \right)
 &\sim \frac 1{|x- y|^2+ 2\Delta^2 -i\varepsilon}\, \d\cF^{ab}( {x},{y}) 
 \left|^x_{y} \right) \nn\\
 \d\cF^{ab}( {x},{y}) &=  \cF^{ab}( {x})-  \cF^{ab}({y}) \ 
 \label{delta-Theta-def}
\end{align}
where $|x-y|$ is the distance in target space $(\R^{9,1},\eta)$,
and $2\Delta^2 \sim L_\NC^2$ is the uncertainty scale on $\cM$.
We can therefore  evaluate the 1-loop integral \eq{Gamma-IKKT-4}
using \eq{trace-coherent-End-hermit}
 as follows
\begin{align}
\Gamma_{\!\textrm{1loop};4}[T]\!
 &=  \frac i4 
  \int\limits_{\cM\times\cM} \frac{\Omega_x \Omega_y}{(2\pi)^m}
  \frac{ \d\cF^{a_1b_1}( {x},{y})
  \d\cF^{a_2b_2}( {x},{y}) \d\cF^{a_3b_3}( {x},{y})\d\cF^{a_4b_4}( {x},{y})}
  {(|{x}- y|^2+2\Delta^2 -i\varepsilon )^4}  \nn\\
 &\qquad \qquad  3\big(-4 \eta_{b_1 a_2} \eta_{b_2 a_3} \eta_{b_3 a_4} \eta_{b_4 a_1}
  + \eta_{b_1 a_2} \eta_{b_2 a_1} \eta_{b_3 a_4} \eta_{b_4 a_3} \Big)  \nn\\[1ex]
  &=  \frac {3i}4 
  \int\limits_{\cM\times\cM}\frac{\Omega_x \Omega_y}{(2\pi)^m}
   \frac{V_4[ \d\cF({ x},{ y})] }{(|x-y|^2+2\Delta^2 -i\varepsilon)^4}
\label{1-loop-coh}
\end{align}
where
\begin{align}
 V_4[ \d\cF] = -4\tr (\d\cF^4) + (\tr \d\cF^2)^2 \ .
 \label{S-4-short}
\end{align}
The  pre-factor $i$ is an artefact as $\Gamma_{\!\textrm{1loop};4}[T]$ 
will pick up an extra $i$ upon evaluating the $i\varepsilon$ regularization.
Note that $ \frac{1}{(|x-y|^2+2\Delta^2 -i\varepsilon)^4}$ is essentially
the  massless Feynman propagator in $9+1$ dimensions, 
for distances larger than the scale of noncommutativity.
Here the maximal SUSY cancellations are crucial, which lead to a residual
interaction which decays like $|x-y|^{-8}$ as appropriate for a massless field 
in $9+1$ dimensions.
This one-loop induced action can indeed be interpreted in terms of IIB supergravity
in $\R^{9,1}$. However it arises only as effective interaction
of objects on the brane $\cM^{3,1} \subset \R^{9,1}$, 
and all the physical dof live on $\cM$ for weak coupling.
Then the $r^{-8}$ behavior thus amounts to a weak, short-distance interaction,
which will be dominated by the long range gravity arising from massless dof on 
$\cM^{3,1}$. 

Note that the double integral in \eq{1-loop-coh}
 is already the explicit bi-local
 {\em 1-loop effective action in position space} on $\cM$. 
This way of computing loops  is very remarkable and has no analog in
the context of ordinary QFT, where no analog of string modes exists.
We will see using a refined treatment  below that its local limit includes the Einstein-Hilbert action for the effective metric on $\cM^{3,1}$,
however only in the presence of fuzzy extra dimensions, i.e. for 
a background brane with the structure 
$\cM^{3,1} \times \cK \subset \R^{9,1}$. Without $\cK$, the local action turns 
out to be a higher-derivative action distinct from ordinary gravity.

\section{Higher-derivative local action on basic branes}
\label{sec:ind-grav-higerderiv}

We start with the simplest possible noncommutative background
\begin{align}
 T^\a = X^\a \sim x^\a:\quad \cM \hookrightarrow \R^{9,1}
\end{align}
for $\a=0,...,3$
 describing a (''basic``) quantized 4-dimensional symplectic space embedded in target space,
with $T^a=0$ for $a=4,...,9$.
Then
\begin{align}
  i[X^\a,X^\b] = \cF^{\a\b}
\end{align}
reduces in the semi-classical regime to the Poisson tensor on $\cM$,
which may  be $x$ -- dependent.
The leading contribution $\Gamma_{\!\textrm{1loop};4}[T]$
\eq{1-loop-coh} to the 1-loop effective action on this background is then given
in terms of the quartic term \eq{S-4-short} 
\begin{align}
   V_4[ \d\cF] &= -4 \tr \d\cF^4 + (\tr \d\cF^2)^2 \nn\\
  &= -4\big(\d\cF^{\a\b} \d\cF_{\b\g}\d\cF^{\g\d}\d\cF_{\d\a} \big)
   + \big(\d\cF^{\a\b} \d\cF_{\a\b}\big)\big(\d\cF^{\g\d} \d\cF_{\g\d}\big) \ .
 \label{V4-explicit}
\end{align}
Here  indices are contracted with $\eta_{\a\b}$, and \eq{delta-Theta-def}
\begin{align}
 \d \cF^{\a\b} = \cF^{\a\b}(x) - \cF^{\a\b}(y) \ .
\end{align}
This action takes the form
\begin{align}
\Gamma_{\rm 1 loop,4}[\cM] = \frac {3i}4\Tr\Big(\frac{V_{4}}{(\Box-i\varepsilon)^4}\Big)
  =  \frac{3i}{4(2\pi)^{4}}\int\limits_{\cM\times \cM}\!\! \Omega_x \Omega_y\,
  \frac{O(\cF(x) - \cF(y))^4}
  {((x-y)^2  - i \varepsilon)^4}\
\label{nonlocal-ind-grav}
\end{align}
representing the contribution of string modes linking $x$ to $y$.
This is a non-local interaction, which is clearly quite different from 4-dimensional gravity. 
We can elaborate its local limit
using the trace formula \eq{trace-NC-conv-int} in terms of the
semi-classical wave-packets $\psi_k(y) =\psi^{(L)}_{k,y}$.
Then the numerator
leads  to 4-th order derivative terms such as
\begin{align}
 \{\cF^{\a\b},\{\cF_{\a\b},\{\cF^{\g\d},\{\cF_{\g\d},\psi_k\}\}\}\}
  &\sim \{\cF^{\a\b},y^\mu\}\{\cF_{\a\b},y^\nu\}
  \{\cF^{\g\d},y^\r\}\{\cF_{\g\d},y^\s\}
  \del_\mu\del_\nu\del_\r \del_\s\psi_k \nn\\
   &=  T^{\a\b\mu} \tensor{T}{_\a_\b^\nu} \tensor{T}{^\g^\d^\r} \tensor{T}{_\g_\d^\s}
  \del_\mu\del_\nu\del_\r\del_\s\psi_k \nn\\
   &\approx  T^{\a\b\mu} \tensor{T}{_\a_\b^\nu} \tensor{T}{^\g^\d^\r} \tensor{T}{_\g_\d^\s}
  k_\mu k_\nu k_\r k_\s\psi_k \
  \label{4th-order-torsion-term}
\end{align}
assuming that $\cF^{\a\b}$ is slowly varying.
 Here we use
the relation with the torsion tensor \eq{torsion-general}
\begin{align}
  \{\cF^{\a\b},y^\mu\} = \tensor{T}{^\a^\b^\mu} \
\end{align}
in local coordinates.
Similarly, the full 4-- derivative contribution of \eq{V4-explicit} takes the form
\begin{align}
 V_{4,\cM}[\d \cF]
   &\approx \Big(\tensor{T}{^\a^\b^\mu} \tensor{T}{_\a_\b^{\nu}}  \tensor{T}{^\g^\d^\r} \tensor{T}{_\g_\d^{\s}}
   - 4 \tensor{T}{^\a^\b^\mu}\tensor{T}{_\b_\g^\n}
   \tensor{T}{^\g^\d^\r} \tensor{T}{_\d_\a^\s}\Big) k_\mu k_\n k_\r k_\s\psi_{k;y}\nn\\
    &=: V_{4}(T)^{\mu\nu\r\s} k_\mu k_\n k_\r k_\s \, \psi_{k;y} \ .
     \label{4th-order-torsion-term-2}
\end{align}
Hence the local contribution to the effective action is
\begin{align}
\Gamma_{\rm 1 loop,4}[\cM] 
 &= \frac{3i}{4(2\pi)^4}\int\limits_{\cM}d^4 x \int d^4 k \,
   \frac{V_{4}(T)^{\mu\nu\r\s} k_\mu k_\n k_\r k_\s}
     {(k\cdot k -i\varepsilon)^4} \
\end{align}
where
 \begin{align}
k\cdot k \equiv k_\mu k_\nu \g^{\mu\nu}  = \r^2  G^{\mu\nu} k_\mu k_\nu
\end{align}
involves the effective metric \eq{eff-gamma-1},
evaluating $\Box \sim \Box_\g$ in local coordinates with $\del\g=0$ using \eq{Box-gamma-psi}.
Being a contraction of 4 torsion tensors,
 this amounts to a higher-derivative contribution
 to the local  action.
 
 At face value, the integral over $k$ is logarithmically 
 divergent both in the IR and the UV,
 but both are artefacts of the present expansion.
 In the IR, the $k$ integration  is effectively 
cut off at the background curvature scale, where \eq{4th-order-torsion-term}
is no longer valid.
This can be  taken into account by adding a suitable mass term 
to the propagator.
In the UV, the loop integral \eq{nonlocal-ind-grav} is finite, 
due to the $\frac{1}{((x-y)^2 - i \varepsilon)^4}$ behavior for long strings.
Hence the above result gives a sub-leading higher-derivative 
contribution to the gravitational action,
which will be dropped. It
can be interpreted as IIB supergravity interaction on $\cM$
due to the exchange of
9+1-dimensional gravitons (and other modes) coupling to the
''matrix`` energy-momentum tensors $\cT_{\a\b}$, which
decays like $(x-y)^{-8}$.

\section{Einstein-Hilbert action from fuzzy extra dimensions}
\label{sec:grav-extra-dim}

Now consider a background with structure $\cM^{3,1} \times \cK$, where 
$\cM^{3,1}$ plays the role of spacetime and
$\cK$ are fuzzy extra dimensions.
Then the one loop effective action in the IKKT model contains a
term which can be interpreted either as  supergravity interaction
between $\cK$ and $\cM$,  or as 
 Einstein-Hilbert action on $\cM$. The effective
Newton constant will be determined by the scale of $\cK$.
Such backgrounds are
realized by matrix configurations where the first 3+1 matrices
\begin{align}
 T^\a  \sim Y^\a :\quad \cM \hookrightarrow \R^{9,1}, \qquad \a =0,...,3 \
\end{align}
describe a quantized 4-dimensional symplectic space $\cM$ as before,
while the remaining 6 matrices
\begin{align}
 T^i \sim m_\cK\cK^i  :\quad \cK \hookrightarrow \R^{9,1} \qquad i=4,...,9 \
\label{fuzzy-extra-dim-BG}
\end{align}
describe $\cK$.
The precise form of $\cK$ is not important
in the following.
The Hilbert space  factorizes accordingly
as $\cH = \cH_\cM \otimes \cH_\cK$,
where $\cH_\cK\cong \C^n$ corresponds to the compact quantum space $\cK$.
Then the trace factorizes  as
\begin{align}
 \Tr_{\Mat(\cH)} &=  \Tr_{\cM} \times  \Tr_{\cK}
\end{align}
using the short notation
\begin{align}
 \Tr_\cM :=  \Tr_{\Mat(\cH_\cM)}, \qquad
 \Tr_\cK :=  \Tr_{\Mat(\cH_\cK)} \ .
\end{align}
The matrix d'Alembertian then decomposes as
\begin{align}
 \Box = [T^\a ,[T_\a ,.]] + [T^i,[T_i,.]]
 = \Box_\cM + \Box_\cK \ .
\end{align}
Let $ \Upsilon_{\L} \in \Mat(\cH_\cK)$ be the eigenmodes of
\begin{align}
 \Box_\cK \Upsilon_{\L} = m^2_\L\, \Upsilon_{\L} \ , \qquad m_{\L}^2 = m_\cK^2 \mu^2_{\L} \ 
 \label{KK-mass}
\end{align}
with dimensionless $\mu^2_{\L}$ and scale $m_\cK$.
We then expand any  $\phi\in \Mat(\cH)$
into a sum of modes
\begin{align}
 \phi_{\L} = \phi_{\L}(y) \Upsilon_{\L}
 \label{phi-product-extra}
\end{align}
with
\begin{align}
 \Box \phi_{\L} = (\Box_\cM + m_{\L}^2) \phi_{\L} \ .
\end{align}
Therefore the $\phi_{\L}$ mode acquires a mass $m_{\L}^2$ on $\cM$,
just like in standard Kaluza-Klein compactification.
 Then $\Tr_\cK$ can be evaluated  by summing over
the ON basis $\Upsilon_{\L}$, while  $\Tr_\cM$ is evaluated
by integrating over the string modes as before.
The key for the following consideration is then  the following decomposition of the potential
\eq{V4-explicit} into contributions from $\cM$, from $\cK$, and mixed
contributions:
 \begin{align}
 V_4[ \d\cF]
  &=  \big(\d\cF^{\a\b} \d\cF_{\a\b}\big)\big(\d\cF^{\g\d} \d\cF_{\g\d}\big)
  -4\big(\d\cF^{\a\b} \d\cF_{\b\g}\d\cF^{\g\d}\d\cF_{\d\a} \big)
    \nn\\
   &\quad + \big(\d\cF^{ij} \d\cF_{ij}\big)
    \big(\d\cF^{kl} \d\cF_{kl}\big)
   -4\big(\d\cF^{ij} \d\cF_{jk}\d\cF^{kl}\d\cF_{li} \big)  \nn\\
  &\quad  + 2\big(\d\cF^{\a\b} \d\cF_{\a\b}\big)
     \big(\d\cF^{ij} \d\cF_{ij}\big) \ .
 \label{S4-explicit-M-K}
\end{align}
There are no other terms, since the mixed commutators
$[T^\a,\cK^i] =0$ vanish due to the
product structure of the background.
The contractions are understood with $\eta_{\a\b}$
or $\d_{ij}$.

\subsection{$\cK$ contribution and effective potential}

Consider the contribution of the $\cK$ terms
\begin{align}
  V_{4,\cK} =  \big(\d\cF^{ij} \d\cF_{ij}\big)
    \big(\d\cF^{kl} \d\cF_{kl}\big)
   -4\big(\d\cF^{ij} \d\cF_{jk}\d\cF^{kl}\d\cF_{li} \big) \ .
\end{align}
Going back to \eq{Gamma-IKKT-4} and summing over the modes \eq{phi-product-extra},
$\d\cF^{ij}$ indicates the term
\begin{align}
\d\cF^{ij} =
 [\cF^{ij},\phi_{\L}(y) \Upsilon_{\L}]
  &= \phi_{\L}(y) [\cF^{ij},\Upsilon_{\L}]
\end{align}
so that $V_{4,\cK}$ is a 4-th order derivation acting on $\Mat(\cH_\cK)$.
For simplicity, we
 assume that $\Upsilon_{\L}$ is a common eigenvector of both $\Box_\cK$
and $V_{4,\cK} $; if not, the following consideration can be  adapted.
Then $V_{4,\cK}$ reduces to
\begin{align}
 \Big([\cF^{ij},[\cF_{ij},[\cF^{kl},[\cF_{kl},.]]]]
   -4[\cF^{ij},[\cF_{jk},[\cF^{kl},[\cF_{li},.]]]] \Big) \Upsilon_{\L}
  =: m_\cK^8 V_{4,\L} \Upsilon_{\L} \
\end{align}
where $m_\cK$ is the scale parameter of $\cK$.
Thus the trace reduces to
\begin{align}
 \Tr\Big(\frac{V_{4,\cK}}{(\Box -i\varepsilon)^4}\Big)
  = m_\cK^8 \sum_{\L} V_{4,\L} \Tr_{\cM} \Big(\frac{1}{(\Box_\cM + m^2_{\L}-i\varepsilon)^4}\Big) \ .
\end{align}
The trace over $\Mat(\cM)$ can be evaluated in the string basis as above,
\begin{align}
  \Tr_{\cM} \Big(\frac{1}{(\Box_\cM + m^2_{\L}-i\varepsilon )^4}\Big)
  &\approx   \frac{1}{(2\pi)^{4}}\, \int\limits_{\cM\times \cM}\!\! \Omega_x \Omega_y\frac{1}{((x-y)^2 + m^2_{\L}-i\varepsilon)^4} \ .
  \label{trace-M-double-K}
\end{align}
The $y$ integration is a convergent short-range integral for $\dim\cM=4$, so that the
discussion for UV-convergent traces on quantum spaces in section \ref{sec:semi-class-Gauss} applies.
We can thus evaluate it as
\begin{align}
  \Tr_{\cM} \Big(\frac{1}{(\Box_\cM + m^2_{\L}-i\varepsilon)^4}\Big)
  &\approx  \frac{1}{(2\pi)^{4}}\,\int\limits_{\cM} d^4x \int d^4 k
  \langle \Psi^{(L)}_{k,x}, \frac{1}{(\Box_\cM + m^2_{\L}-i\varepsilon)^4}
  \Psi^{(L)}_{k,x}\rangle \nn\\
   &\approx  \frac{1}{(2\pi)^{4}}\,\int\limits_{\cM} d^4x \sqrt{G}
  \int \frac{d^4 k}{\sqrt{G}}
   \frac{1}{(k\cdot k + m_{\L}^2-i\varepsilon)^4}  \
   \label{trace-M-k}
\end{align}
  using \eq{trace-NC-conv-int}, and assuming the normalization
$\langle \Psi^{(L)}_{k,x}, \Psi^{(L)}_{k,x}\rangle = 1$.
This boils down to
\begin{align}
 \k_{(4)} := \int \frac{d^4 k}{\sqrt{G}}\,\frac{1}{(k\cdot k + m^2 - i \varepsilon)^4} \
 = i\frac{\pi^2}{6 m^4} \r^{-4}
\end{align}
 using \eq{kappa-3-4-explicit}, where
$k\cdot k \equiv \r^2  G^{\mu\nu} k_\mu k_\nu$.
Thus the contribution of $\cK$ to the effective action is
\begin{align}
 \Gamma_{\rm 1 loop}^\cK &=
 \frac{3i}4\Tr\Big(\frac{V_{4,\cK}}{(\Box - i \varepsilon)^4}\Big)
  = -\frac{\pi^2}{8(2\pi)^{4}}
  \int\limits_\cM d^4 x\sqrt{G}\, \r^{-4} m_\cK^4\, \big(\sum_{\L} \frac{1}{\mu^4_{\L}} V_{4,\L}\big) \nn\\
   &=-\frac{1}{8(4\pi)^{2}}
  \int\limits_\cM \Omega\,\r^{-2} m_\cK^4\, \big(\sum_{\L} \frac{1}{\mu^4_{\L}} V_{4,\L}\big)   \ 
  \label{1-loop-pot-cK}
\end{align}
where
\begin{align}
 d^4 x \r_M = \Omega = \sqrt{G}\, \r^{-2}
\end{align}
is the symplectic volume form on $\cM^{3,1}$, and  $\mu^2_{\L}$ are
the dimensionless KK masses \eq{KK-mass}.
This term contributes to the effective potential
for the  $\cK^i$ which form the extra dimension $\cK$.
Note that the integral measure is given by the
symplectic volume form $\Omega = d^4 x \rho_M$ rather than the Riemannian one,
with important
implications for the cosmological constant problem.

\subsection{Mixed contributions and induced gravity action}
\label{sec:mix-grav-basic}

The most interesting contribution comes from the mixed term in \eq{S4-explicit-M-K}
\begin{align}
  V_{4,\rm mix} =   2\big(\d\cF^{\a\b} \d\cF_{\a\b}\big)
     \big(\d\cF^{ij} \d\cF_{ij}\big) \ .
     \label{V4-mixed-K}
\end{align}
As before
we expand all modes into product states of the form
 $\phi_{\L} = \phi_{\L}(y) \Upsilon_{\L}$;
 then the first factor acts only on $\phi_{\L}(y)$, and the second  only on
 $\Upsilon_{\L}$.
 We assume again  that
  $\Upsilon_{\L}$ is a common eigenvector of both $\Box_\cK$
and the 2nd order derivation
$\big(\d\cF^{ij} \d\cF_{ij}\big)$ acting on $\Mat(\cH_\cK)$.
Then
\begin{align}
\big(\d\cF^{ij} \d\cF_{ij}\big)\Upsilon_{\L} =
 [\cF^{ij} ,[\cF_{ij},\Upsilon_{\L}]] = m_\cK^4 C^2_{\L} \Upsilon_{\L} \ ,
 \qquad C^2_{\L} >0 \
 \label{K-mixed-contrib}
\end{align}
where the $C^2_{\L} > 0$ depend on the structure of $\cK$,
but are positive
since the corresponding target space metric is Euclidean.

Next, we need to evaluate $\big(\d\cF^{\a\b} \d\cF_{\a\b}\big)$
acting on $\Mat(\cH_\cM)$.
Using the string representation,
the trace  $\Tr_\cM\big(\frac{V_{4,{\rm mix}}}{\Box^4}\big)$
leads again to a non-local integral $\int_{\cM\times\cM} \Omega_x \Omega_y$ as in \eq{trace-M-double-K}, which is convergent for large $|x-y|$ and hence almost-local.
We  compute it again using the
trace formula \eq{trace-NC-conv-int}
in terms of the  wave packets $\Psi^{(L)}_{k;y} \sim \psi^{(L)}_{k;y}(x)$
with size $L$, which we choose somewhat shorter than the typical scale of the background 
geometry.
For the numerator, we recall the formula \eq{VF-wavepackets}
\begin{align}
 v[\psi_{k;y}] =
v^\mu\del_\mu\psi_{k;y} &\approx i k_\mu v^\mu \psi_{k;y} \ ,
\qquad k > \frac 1L \ .
\end{align}
This allows to evaluate the 2nd order differential operator
$\d\cF^{\a\b} \d\cF^{\a\b} \sim - \{\cF^{\a\b}, \{\cF^{\a\b},.\}\}$ acting on
the wave-packets $\psi_{k;y}$ as
\begin{align}
 \{\cF^{\a\b},\psi_{k;y}\} &\approx i k_\mu \{\cF^{\a\b},x^\mu\} \psi_{k;y}
  = i k_\mu \tensor{T}{^\a^\b^\mu}  \psi_{k;y}  \nn\\
 \{\cF^{\a\b}, \{\cF_{\a\b},\psi_{k;y}\}\}
 &\approx i k_\mu \{\cF^{\a\b},\tensor{T}{_\a_\b^\mu}  \psi_{k;y} \}  \nn\\
  &= i k_\mu \{\cF^{\a\b},\tensor{T}{_\a_\b^\mu}  \} \psi_{k;y}
  + i k_\mu\tensor{T}{^\a^\b^\mu}  \{\cF_{\a\b}, \psi_{k;y} \}\nn\\
  &= i k_\mu  \{\cF^{\a\b},\tensor{T}{_\a_\b^\mu}  \}  \psi_{k;y}
  - \tensor{T}{^\a^\b^\mu} \{\cF_{\a\b}, x^{\nu}  \} k_\mu k_{\nu}\psi_{k;y} \nn\\
  &\approx -\tensor{T}{^\a^\b^\mu} \tensor{T}{_\a_\b^{\nu}}\, k_\mu  k_{\nu}\psi_{k;y}\ ,
   \qquad k > \frac 1L \ .
   \label{derivations-spacetime-torsion}
\end{align}
We  neglect the first term assuming that the curvature scale is smaller than $k^2$.
Putting this together,
 $V_{4,\rm mix}$ acting on  $\psi_{k;y}\Upsilon_{\L}$ is
 approximately diagonal, and reduces to
\begin{align}
 V_{4{\rm mix}} [\psi_{k;y}\Upsilon_{\L}]
 = 2 m_\cK^4 C^2_{\L} \tensor{T}{^\a^\b^\mu}  \tensor{T}{_\a_\b^{\nu}}\, k_\mu k_{\nu}
 \ \psi_{k;y}  \Upsilon_{\L} \ .
\end{align}
Hence the full trace reduces using \eq{trace-NC-conv-int} to the following local integral
\begin{align}
 \Tr\Big(\frac{V_{4,{\rm mix}}}{(\Box-i\varepsilon)^4}\Big)
   &= \frac{2m_\cK^4}{(2\pi)^4} \int\limits_\cM d^4 x\sqrt{G}\, \tensor{T}{^\a^\b^{\mu}}\tensor{T}{_\a_\b^{\nu}}
  \sum_{\L} C^2_{\L} \int \frac{d^4 k}{\sqrt{G}}\, \frac{k_\mu  k_{\nu}}
  {(k\cdot k + m^2_{\L} - i \varepsilon)^4}  \nn\\
   &= \frac{2m_\cK^4}{(2\pi)^4}\int\limits_\cM\,d^4 x\sqrt{G}\,
     \big(\sum_{\L} C^2_{\L} \tilde\k_{(4)}(m_{\L})\big) \tensor{T}{^\a^\b^{\mu}}\tensor{T}{_\a_\b^{\nu}} G_{\mu\nu} \ .
   \label{tr-M-mix}
\end{align}
 As before $k\cdot k = \g^{\mu\nu} k_\mu k_\nu$ is defined via 
 the effective metric without dilaton,
 and
\begin{align}
  \tilde\k_{(4)} G_{\mu\nu}
  &= \int \frac{d^4 k}{\sqrt{G}}\, \frac{k_\mu  k_{\nu}}{(k\cdot k + m^2 - i \varepsilon)^4} \
\end{align}
where $\tilde\k_{(4)} = \tilde\k_{(4)}(m)$ can be computed by taking the trace
\begin{align}
 4\r^2\tilde\k_{(4)} + m^2 \k_{(4)} &= \int \frac{d^4 k}{\sqrt{G}}
 \,\frac{k\cdot k + m^2}{(k\cdot k  + m^2 - i \varepsilon)^4} \nn\\
  &= \int \frac{d^4 k}{\sqrt{G}}\,\frac{1}{(k\cdot k  + m^2 - i \varepsilon)^3} = \k_{(3)}
  \label{tilde-kappa-4}
\end{align}
in terms of $\k_{(3)}$ and $\k_{(4)}$ given in \eq{kappa-n}.
This gives
\begin{align}
 4\r^2\tilde\k_{(4)}(m) 
 =  \frac{i\pi^2}{3m^2} \r^{-4} \ .
 \label{kappa-4-tilde}
 \end{align}
 Now recall that  the mass scale of the KK modes is set by  $m_\cK$.
Using the conventions that the indices of the torsion tensor are
covariantized in terms of the frame $E^{\dot\a}_\mu$ corresponding to the
metric $\g^{\mu\nu} = \r^2 G^{\mu\nu}$ \cite{Steinacker:2020xph}, we can rewrite
\begin{align}
 \r^{-2} \tensor{T}{^{\dot\a}^{\dot\b}^{\mu}}\tensor{T}{_{\dot\a}_{\dot\b}^{\nu}} G_{\mu\nu}
   &=  \tensor{T}{_\r_\s^{\mu}}\tensor{T}{_{\r'}_{\s'}^{\nu}} \g^{\r\r'} \g^{\s\s'} \g_{\mu\nu}
    = \tensor{T}{^\r_\s_{\mu}}\tensor{T}{_{\r}^{\s}_{\mu}} \g^{\mu\mu'}
    = \r^2 \tensor{T}{^\r_\s_{\mu}}\tensor{T}{_{\r}^{\s}_{\mu}} G^{\mu\mu'} \ .
    \label{torsion-contraction-frame-G}
\end{align}
Collecting these results,
we obtain
\begin{align}
 \Gamma_{\rm 1 loop}^{\cK-\cM}  
 =\frac{3i}4\Tr\Big(\frac{V_{4,{\rm mix}}}{(\Box-i\varepsilon)^4}\Big)
  &= -\frac{1  }{(2\pi)^4}\int\limits_\cM d^4 x\sqrt{G}\, \r^{-2} m_\cK^2\,c^2_{\cK} \,
   \tensor{T}{^\r_\s_{\mu}}\tensor{T}{_{\r}^{\s}_{\nu}} G^{\mu\nu} \nn\\
 &= - \frac 12\int\limits_\cM d^4 x  \,\frac{\sqrt{G}}{16\pi G_N}
  \tensor{T}{^\r_\s_{\mu}}\tensor{T}{_{\r}^{\s}_{\nu}} G^{\mu\nu}\ . 
   \label{Gamma-EH-0}
\end{align}
Here
 \begin{align}
 c_{\cK}^2 = \frac{\pi^2}{8}\sum_{\L} \frac{C^2_{\L}}{\mu_{\L}^2}\  > 0 \ 
 \label{C2-K-cutoff}
\end{align}
is finite  due to the fuzzy nature of $\cK$, and 
\begin{align}
  \frac{1}{G_N} = \frac{2c^2_{\cK}}{\pi^3} \r^{-2} m_\cK^2\,
  \label{Newton-constant-rho-mK}
 \end{align}
 plays the role of the Newton constant i.e. the Planck scale, which is set by
 the compactification scale via $\r^{-2}m_\cK^2$.
 This can be rewritten using
the identity \eq{R-offshell-T-2} as
\begin{align}
\boxed{\
 \Gamma_{\rm 1 loop}^{\cK-\cM}
 =  \int\limits_\cM \! d^{4}x\frac{\sqrt{|G|}}{16 \pi G_N}\,
   \Big(\cR
  + \frac 12\tilde T_{\nu} \tilde T_{\mu}  G^{\mu\nu}  
        - 2 \r^{-2} \del_\mu\r\del^\mu\r
        + 2 \r^{-1} \del_\mu\r\, G_N^{-1}\del^\mu G_N \Big) \ .
 }
   \label{Gamma-EH-I}
\end{align}
Here $\tilde T_{\mu}$ encodes the  totally antisymmetric part of the torsion
\eq{T-AS-dual-2}, which reduces to the axion using the
on-shell relation $\tilde T_\mu = \r^{-2}\del_\mu\tilde\rho$ \eq{T-del-onshell}.
Hence for  $\tilde T_\mu =0 = \del\rho$ we recover
the Einstein-Hilbert action $\int G_N^{-1}\cR + \cL_{\rm matter}$  coupled to matter
leading to (modified) Einstein equations  with the appropriate sign, since
our sign convention entails
$\frac{\d S_{\rm matter}}{\d G^{\mu\nu}} = -  T_{\mu\nu}$.

 It is gratifying to see that the Planck scale is related to the Kaluza-Klein scale $m^2_\cK$ for the
 fuzzy extra dimensions $\cK$, and hence to the  UV cutoff on $\cK$.
The effective Newton constant $G_N$ is hence determined dynamically in terms of the dilaton and $m_\cK^2$. The stabilization of the latter will be discussed  in section \ref{sec:1-loop-grav-stab}.
The presence of the dilaton  may seem reminiscent of Brans-Dicke theory,
however $\r$ is not an independent field here but may be determined by other constraints.

Before we proceed, let us recapitulate this important result:
the one loop effective action for $\cM^{3,1}\times\cK$ leads to a
gravitational action on $\cM^{3,1}$ including the Einstein-Hilbert term.
On the other hand, this induced action can also be interpreted as IIB supergravity interaction
between $\cK$ and $\cM$ in target space $\R^{9,1}$.
This term can hence be considered as {\bf almost-local}  IIB interaction.
In other words, {\em the local 3+1-dimensional gravity} {\bf  action} arises from the
{\em non-local IIB supergravity} {\bf interaction} in 9+1 dimensions.
Since all fluctuations propagate on the space-time brane $\cM$ in the weak coupling regime,
there is no need to compactify target space (unlike in string theory), and
no landscape problem arises.

\subsection{$\cM$ contributions revisited}

Finally, we reconsider the contribution of the terms
\begin{align}
  V_{4,\cM} =  \big(\d\cF^{\a\b} \d\cF^{\a\b}\big)
    \big(\d\cF^{\g\d} \d\cF^{\g\d}\big)
   -4\big(\d\cF^{\a\b} \d\cF^{\b\g}\d\cF^{\g\d}\d\cF^{\d\a} \big) \
\end{align}
arising from $\cM$, in the presence of $\cK$.
Summing  over the modes \eq{phi-product-extra},
we can evaluate the trace over $\Mat(\cM)$
as in section \ref{sec:ind-grav-higerderiv} by integrating over the string modes.
The 4th order differential operator acting on
the wave-packets $\psi_{k;y}$ leads again to
\begin{align}
 V_{4,\cM}[\psi_{k;y}]
    &\approx V_{4}(T)^{\mu\nu\r\s}k_\mu k_\n k_\r k_\s
\end{align}
as in \eq{4th-order-torsion-term-2}. All KK modes on $\cK$ contribute in this way,
and their  mass enters in the propagator.
Therefore the local contribution to the effective action takes the form
\begin{align}
 i\Tr\Big(\frac{V_{4,\cM}}{(\Box-i \varepsilon)^4}\Big)
 &= \frac{i}{(2\pi)^4}\int\limits_{\cM}d^4 x \int d^4 k
   \sum_{\L}\,\frac{V_{4}(T)^{\mu\nu\r\s} k_\mu k_\n k_\r k_\s}
     {(k\cdot k + m^2_{\L}-i\varepsilon)^4} \
     \label{nonlocal-ind-grav-K}
\end{align}
summing over the KK modes on $\cK$.
The integral over $k$ is again log-divergent as in
section \ref{sec:ind-grav-higerderiv}, but regulated by the
fully non-local form \eq{nonlocal-ind-grav} of the induced action.
This is  expected to be a sub-leading higher-derivative contribution 
to 4-dimensional gravity, which we will not consider any further.

\section{Gravity on covariant quantum spacetime}
\label{sec:ind-grav-covar}

Now we want to compute the one loop effective action on a covariant $\cM^{3,1}$ background
given by some deformation of the FLRW space-time discussed in \cite{Sperling:2019xar}.
This is a quantization of a 6-dimensional symplectic space which is a $S^2$
 bundle over space-time,
\begin{align}
 \C P^{1,2} \cong S^2_n \ \bar\times\ \cM^{3,1} \ .
\end{align}
Here $\bar\times$ indicates the twisted (=equivariant) bundle structure.
 The space of functions is then spanned (locally) by the following
 modes
\begin{align}
 \phi_{sm} = \phi_{sm}(y) Y_{sm}(t) \ .
 \label{modes-harmonics-S2n}
\end{align}
Here $y$ are coordinates on  $\cM^{3,1}$, and $Y_{sm}(t)$ are (normalized) fuzzy spherical harmonics on $S^2_n$ corresponding
to the expansion into $\hs$ modes \cite{Sperling:2019xar}.
These are eigenvectors of $\Box$
with eigenvalue\footnote{Here we anticipate that the trace is UV finite, so that only semi-classical modes contribute.
We can then use the trace formula \eq{trace-wavelets} for the (semi-) classical case, which coincides with the one
in the quantum case as  discussed in section \ref{sec:semi-class-Gauss}.}
\begin{align}
 \Box = k\cdot k  +  m^2_s \ .
 \label{Box-EV-covar}
\end{align}
Here $k\cdot k = k_\mu k_\nu \g^{\mu\nu}$ where
\begin{align}
 \g^{\mu\nu} = \eta_{\a\b} E^{\a\mu} E^{\b\nu} \
\end{align}
describes a 
FLRW space-time with effective metric $G_{\mu\nu} = \r^{2} \g_{\mu\nu}$ 
with scale factor $a(t) \sim t$ at late times, which arises from the 
effective frame $E^{\a\mu}$ on $\cM^{3,1}$.
The ''higher-spin mass`` on $S^2$ is given by 
\begin{align}
 m^2_s = \frac{3s}{R^2} \  
\end{align}
which  is an  IR mass in the late-time regime.

The desired local contribution to the one loop effective action can now be computed in the semi-classical setting.
It is crucial here that the internal space $S^2_n$ supports only finitely many modes.
We will therefore evaluate the trace over $\Mat(\cH)$ in a hybrid way, expanding into
the finite tower of $\hs$ harmonics on the internal $S^2_n$, and subsequently using the
semi-classical trace formula for localized wavepackets on $\cM^{3,1}$ as above.
Since the Poisson structure on the bundle space does not respect the local factorization,
there will  be mixed $S^2-\cM$ contributions
 with a  non-standard  behavior.
However these are suppressed in the
presence of fuzzy extra dimensions, which span the transversal 6 dimensions in
target space as above. Only then 
a physically acceptable gravity action will arise, as
elaborated in section \ref{sec:cov-quant-ind-grav-extra}.

\subsection{Gravitational action on $\cM^{3,1}$ without extra dimensions}
\label{sec:cov-quant-ind-grav}

We start again with the general one loop effective action \eq{Gamma-IKKT},
where now
\begin{align}
 \cF^{\a\b} = -\{t^\a,t^\b\} = \frac 1{r^2 R^2} \theta^{\a\b} \ .
\end{align}
Expanding the modes into normalized harmonics on $S^2_n$ \eq{modes-harmonics-S2n},
each derivative
 $\d\cF^{\a\b} = [\cF^{\a\b},.]$ in
\begin{align}
  V_{4} =  \big(\d\cF^{\a\b} \d\cF_{\a\b}\big)
    \big(\d\cF^{\g\d} \d\cF_{\g\d}\big)
   -4\big(\d\cF^{\a\b} \d\cF_{\b\g}\d\cF^{\g\d}\d\cF_{\d\a} \big) \
\end{align}
acts  either on
$\phi_{sm}(y)$  or on $Y_{sm}(t)$ via
\begin{align}
 [\cF^{\a\b},\phi_{sm}(y) Y_{sm}(t)] =
 [\cF^{\a\b},\phi_{sm}(y)] Y_{sm}(t)
  +\phi_{sm}(y) [\cF^{\a\b}, Y_{sm}(t)] \ .
  \label{derivations-M-S2}
\end{align}
This leads to  contributions from $\cM$, from $S^2_n$, and mixed
$S^2_n - \cM$ contributions, where $S^2_n$ plays a role somewhat analogous to $\cK$.

\subsubsection{Pure $\cM$ contribution}

The pure $\cM$ contribution is obtained by keeping only the first term in \eq{derivations-M-S2}. This is completely analogous
as in section \ref{sec:ind-grav-higerderiv} summing
over the internal harmonics, however
$\Box$ now acts on the non-trivial modes \eq{modes-harmonics-S2n} with  eigenvalue
\begin{align}
 \Box = k\cdot k  + m^2_s \ .
\end{align}
This gives a similar
non-local conribution as in \eq{nonlocal-ind-grav}, which
we will not consider  any further.

\subsubsection{Mixed $S^2 - \cM$ contributions}

Now we take into account contributions from the second term in \eq{derivations-M-S2}.
Since the mixed $S^2 - \cM$ contribution will turn out to be sub-leading compared with those
from $\cK$, we simplify the computation by assuming that the local structure is the same as
on the undeformed cosmic background.
Then $[\cF^{\a\b},t^i] = -\frac i{R^2}(\eta^{\a i}t^\b - \eta^{\b i}t^\a)$,
and
$\Box$ respects the local $SO(3)$.
One then finds locally
\begin{align}
 \sum_m\langle Y_{sm}(t),[\cF^{\a\b},Y_{sm}(t)]\rangle &= 0  \nn\\
 \sum_m\langle Y_{sm}(t),[\cF^{\a\b},[\cF^{\g\d},Y_{sm}(t)]]\rangle
 &= \frac {1}{4R^4} \k^{\a\b;\g\d}_s  \nn\\
 \sum_m\langle Y_{sm}(t),[\cF^{\a\b},[\cF^{\g\d},[\cF^{\a'\b'},[\cF^{\g'\d'},Y_{sm}(t)]]]]\rangle
 &= \frac {1}{4R^8} \k^{\a\b;\g\d;\a'\b';\g'\d'}_s
\end{align}
where the tensors $\k^{\a\b;\g\d}_s$ etc.  are invariant under
the local $SO(3)$ as well as under
the space-like $SO(3,1)$ isometry. 
The remaining derivations $\d\cF^{\a\b}$ act on $\cC(\cM^{3,1})$
as in \eq{derivations-spacetime-torsion}.
Taking into account the form of $V_4$,
one obtains  the following mixed  $S^2 - \cM$ contributions from $V_4$:
\begin{align}
 \sum_m\langle Y_{sm}(t), V_{4,{\rm mix}}Y_{sm}(t)\rangle
 &= \frac 1{2 R^4} \k^{\a\b;\a\b}_s
  \big(\d\cF^{\g\d} \d\cF^{\g\d}\big)
  +  \frac 1{R^4}\k^{\a\b;\g\d}_s\d\cF^{\a\b}\d\cF^{\g\d}\nn\\
  &\quad -  \frac 4{R^4}\k^{\a\b;\b\g}_s\d\cF^{\g\d}\d\cF^{\d\a}
  -  \frac 2{R^4}\k^{\a\b;\g\d}_s\d\cF^{\b\g}\d\cF^{\d\a}   \nn\\
  &=: \frac 2{R^4}\tilde k_{\a\b\g\d}^s\,\d\cF^{\a\b}\d\cF^{\g\d}
\end{align}
where $\d\cF^{\a\b}\d\cF^{\g\d}$ are second order derivations acting on $\cC(\cM^{3,1})$
which needs to be traced over.
This should be compared with \eq{K-mixed-contrib} for fuzzy extra dimensions.
The trace decomposes into a finite sum of higher spin sectors
and the trace over the
scalar functions $\phi_{sm}(y)$. We can evalute the
latter  using the classical trace formula
similarly as in  \eq{tr-M-mix}, which  leads to
\begin{align}
 \Gamma_{\rm 1 loop}^{S^2 - \cM}
 = \frac{3i}4 \Tr\Big(\frac{V_{4,{\rm mix}}}{(\Box-i\varepsilon)^4}\Big)
  &= \frac{3i}{2(2\pi)^4R^4} \int\limits_{\cM} d^4 x \sqrt{G}\,
  \tensor{T}{^\a^\b^\mu}\tensor{T}{^\g^\d^{\nu}} \sum_s\tilde k^s_{\a\b\g\d}
  \int \frac{d^4 k}{\sqrt{G}}
  \frac{ k_\mu  k_{\nu}}{(k\cdot k + m^2_{s}-i\varepsilon)^4} \nn\\
   &= \frac{3i}{2(2\pi)^4} \frac{1}{R^4} \int\limits_\cM d^4 x \sqrt{G}\,
   \big(\sum_s \tilde \k_{(4)}(m_s^2)\tilde k^s_{\a\b\g\d}\big)
   \tensor{T}{^\a^\b^{\mu}}\tensor{T}{^\g^\d^{\nu}} G_{\mu\nu}  \nn\\
&= -\frac{1}{(2\pi)^4} \frac{\pi^2}{8R^2} \int\limits_\cM d^4 x\sqrt{G} \r^{-6}\,
    \Big(\sum_{s>0} \frac {\tilde k^s_{\a\b\g\d}}{3s}\Big)
    \tensor{T}{^\a^\b^{\mu}}\tensor{T}{^\g^\d^{\nu}} G_{\mu\nu} \ ,
   \label{tr-M-S2}
\end{align}
 recalling that \eq{kappa-4-tilde}
\begin{align}
 \tilde\k_{(4)}(m^2_s) = \frac{i\pi^2}{12 m^2_s}\r^{-6} = \frac{i\pi^2 R^2}{36 s} \r^{-6} \ .
 \label{kappa-4-tilde-2}
\end{align}
Note that the contractions of the torsion
via $\tilde k_{\a\b\g\d}$ is not Lorentz invariant a priori, reflecting the presence of
a time-like vector field  on the FLRW space-time.
However,  we will argue  that this is strongly suppressed compared with the contribution from $\cK$ 
considered in the next section provided
\begin{align}
 c_\cK^2 m^2_\cK \gg \frac{1}{R^2} \ .
\end{align}

\subsubsection{Pure $S^2_n$ contribution}

Finally, we consider the pure contribution from $S^2_n$.
The trace then reduces to the form
\begin{align}
\Gamma_{\rm 1 loop}^{S^2} =
 \frac{3i}4\Tr\Big(\frac{V_{4,S^2}}{(\Box - i \varepsilon)^4}\Big)
  &= \frac{3i}{4(2\pi)^4}
  \int\limits_\cM d^4 x\,\sum_{s} \frac{\tilde V_{4,s}}{R^8} \int d^4 k\,
  \frac{1}{(k\cdot k  + m^2_s - i \varepsilon)^4} \nn\\
  &= -\frac{3}{4(2\pi)^4}
  \int\limits_\cM d^4 x\sqrt{G}\,\sum_{s} \frac{\tilde V_{4,s}}{R^8}
  \k_{(4)}(m^2_s)   \nn\\
  &= -\frac{1}{8(4\pi)^2}
  \int\limits_\cM \Omega\, \r^{-2} \frac 1{R^4}\sum\limits_{s\geq 1}
  \frac{\tilde V_{4,s}}{\big(3s\big)^2}  \ 
  \label{S2-cov-pure}
 \end{align}
 for some dimensionless $\tilde V_{4,s}$.
 This amounts to a contribution
to the effective potential, which a priori could have either sign.

\subsection{Einstein-Hilbert action from fuzzy extra dimensions $\cK$}
\label{sec:cov-quant-ind-grav-extra}

Finally, we combine the above considerations and consider the
case of fuzzy extra dimensions $\cK$ on covariant quantum spaces,
so that the full background has the structure
\begin{align}
S^2_n \ \bar\times \ \cM^{3,1} \times\cK \ .
\end{align}
Such a background will typically involve all 9+1 matrices for suitable $\cK$,
thus exploiting the full IKKT matrix model.
The space of functions is then spanned by the following  modes on both $\cK$ and $S^2_n$,
\begin{align}
 \phi_{\L,sn}(y) Y_{sn}(t)\Upsilon_{\L} \
 \label{K-S-M-modes}
\end{align}
where $\Upsilon_{\L}$ are the eigenmodes on $\cK$.
Generalizing \eq{Box-EV-covar}, these are eigenvectors of $\Box$
with eigenvalue
\begin{align}
 \Box = k\cdot k  + m^2_s + m^2_{\L} \ .
\end{align}
We should distingish the contributions from the trivial and non-trivial modes on $\cK$.
The trivial contributions coincide with those of section
\ref{sec:cov-quant-ind-grav} in the
absence of $\cK$.
To simplify the discussion, we will focus on the
contribution from the non-trivial modes in the following.

\subsubsection{Mixed $\cM^{3,1} - \cK$ contribution and induced gravity}

Consider the contributions of the mixed $4D-6D$  terms
\eq{V4-mixed-K}
\begin{align}
  V_{4,\rm mix} =   2\big(\d\cF^{\a\b} \d\cF_{\a\b}\big)
     \big(\d\cF^{ij} \d\cF_{ij}\big) \
     \label{V4-mixed-K-2}
\end{align}
acting on non-trivial modes on  $\cK$ and $\cM^{3,1}$,
\begin{align}
 \phi_{\L}(y) \Upsilon_{\L} \ .
\end{align}
The mixed term from $V_4$  acting on $\cK$ and
functions on $\cM^{3,1}$ have the same structure as in section \eq{sec:mix-grav-basic},
and lead to the same effective  action as in \eq{tr-M-mix} and \eq{Gamma-EH-0},
\begin{align}
 \Gamma_{\rm 1 loop}^{\cK - \cM}
 &= \frac{3i}4 \Tr\Big(\frac{V_{4,{\rm mix}}}{\Box -i\varepsilon)^4}\Big) \nn\\
    &= i\frac{3 }{2(2\pi)^4} \int\limits_\cM d^4 x\, \tensor{T}{^\a^\b^{\mu}}\tensor{T}{^\a^\b^{\nu}}
   \sum_{\L,s} (2s+1)C^2_{\L} m_\cK^4\int d^4 k\, \frac{k_\mu  k_{\nu}}
   {(k\cdot k + m^2_{\L} + m^2_s - i \varepsilon)^4}  \nn\\
  &= -\frac{1 }{(2\pi)^4}\int\limits_\cM d^4 x\sqrt{G}\,\r^{-2} m_\cK^2\,c^2_{\cK}\,
   \tensor{T}{^\r_\s_{\mu}}\tensor{T}{_{\r}^{\s}_{\nu}} G^{\mu\nu}
   \label{tr-M-mix-2-cov}
\end{align}
where $c_{\cK}^2$ \eq{C2-K-cutoff} is replaced by
 \begin{align}
 c_{\cK}^2 = \frac{\pi^2}{8}\sum_{\L,s}
 \frac{(2s+1) C^2_{\L}}{\mu_{\L}^2 + \frac{m_s^2}{m^2_\cK}} \ .
 \label{C2-K-cutoff-cov}
\end{align}
Again, we can rewrite the effective action
in terms of an Einstein-Hilbert term \eq{Gamma-EH-I}
\begin{align}
\boxed{\
 \Gamma_{\rm 1 loop}^{\cK-\cM}
 =  \int\limits_\cM \! d^{4}x\frac{\sqrt{|G|}}{16 \pi G_N}\,
   \Big(\cR
  + \frac 12\tilde T_{\nu} \tilde T_{\mu}  G^{\mu\nu}  
        - 2 \r^{-2} \del_\mu\r\del^\mu\r
        + 2 \r^{-1} \del_\mu\r\, G_N^{-1}\del^\mu G_N \Big) \ .
 }
   \label{Gamma-EH-I-cov}
\end{align}
with effective  Newton constant
\eq{Newton-constant-rho-mK}
\begin{align}
  \frac{1}{G_N} = \frac{2c^2_{\cK}}{\pi^3} \r^{-2} m_\cK^2\ .
  \label{Newton-constant-rho-mK-2}
 \end{align}
Note that this action has a similar structure as the  $S^2 - \cM$  contribution \eq{tr-M-S2},
however the latter is not  Lorentz invariant, and should be suppressed  to obtain a
reasonable gravity theory.
Since \eq{tr-M-mix-2-cov} is governed by
the scale factor $m_\cK^2$ rather than $\frac{1}{R^2}$,
this is the case provided
\begin{align}
 c_\cK^2 m^2_\cK \gg \frac 1{R^2} \ \sim m_s^2
  \label{K-larger-S2}
\end{align}
so that the  contribution from $\cK$ dominates.
This is  reasonable because the Kaluza-Klein
modes arising from $\cK$ should be very heavy, while $m^2_s$ should be a very small IR mass scale.
The scale of $m^2_\cK$ is determined dynamically  by quantum effects, which will be
discussed below.

\subsubsection{Mixed $S^2 - \cK$ contribution}

Now consider contributions of the mixed $4D-6D$ terms
\eq{V4-mixed-K}
\begin{align}
  V_{4,\rm mix} =   2\big(\d\cF^{\a\b} \d\cF_{\a\b}\big)
     \big(\d\cF^{ij} \d\cF_{ij}\big)
     \label{V4-mixed-K-2-2}
\end{align}
where $\d\cF^{\a\b} \d\cF_{\a\b}$
acts on  $Y_{sn}(t)$ rather than $\phi_\L(y)$, and
the transversal derivations act  on $\Upsilon_{\L}$ as in \eq{K-mixed-contrib}, with
\begin{align}
\big(\d\cF^{ij} \d\cF_{ij}\big)\Upsilon_{\L} =
 [\cF^{ij} ,[\cF_{ij},\Upsilon_{\L}]] = m_\cK^4 C^2_{\L} \Upsilon_{\L} \ ,
 \qquad C^2_{\L} >0 \
 \label{K-mixed-contrib-2}
\end{align}
 under the same assumptions.
Now
\begin{align}
 C^2_{\cM} = \d\cF^{\a\b} \d\cF_{\a\b}
\end{align}
 is the quadratic Casimir of
$\mso(3,1)$, which is diagonal for the spherical harmonics on $S^2_n$
with positive eigenvalue
\begin{align}
 C^2_{\cM} Y^s_m &= \frac 2{R^4}\, s(s+2) Y^s_m \ .
\end{align}
As a check, we compute
\begin{align}
C^2_{\cM} t^i &=
 [\cF_{\a\b},[\cF^{\a\b},t^i]]
 = -\frac{i}{R^2}[\cF_{\a\b}, \eta^{\a i} t^\b - \eta^{\b i}t^\a]
  = \frac{6}{R^4} t^i  \ .
\end{align}
Summing over the modes \eq{K-S-M-modes},
$V_{4,\rm mix}$ reduces to
\begin{align}
V_{4,\rm mix} [\Upsilon_{\L} Y_{sn}(t)]
  =  \frac{4 m_\cK^4}{R^4}\, s(s+2) C^2_{\L}\, \Upsilon_{\L} Y_{sn}(t)\
\end{align}
and the trace can be evaluated as
\begin{align}
  \frac{3i}4\Tr\Big(\frac{V_{4,\cK}}{(\Box -i \varepsilon)^4}\Big)
  &= i\frac{3m_\cK^4}{(2\pi)^4 R^4}
  \int\limits_\cM d^4 x\,\sum_{\L;sn} s(s+2) C^2_{\L} \int d^4 k\,
  \frac{1}{(k\cdot k  + m^2_s+ m^2_{\L}-i \varepsilon)^4} \nn\\
  &= i\frac{3m_\cK^4}{(2\pi)^4 R^4}
  \int\limits_\cM d^4 x\sqrt{G}\,\sum_{\L;sn} s(s+2) C^2_{\L}
  \k_{(4)}(m^2_s+ m^2_{\L})   \nn\\
  &= -\frac{\pi^2 }{2(2\pi)^4 }
  \int\limits_\cM \Omega\, \r^{-2}\frac{m_\cK^4}{R^4}\sum_{\L;s}
  \frac{s(s+2)(2s+1) C^2_{\L}}{\big(m^2_s+ m^2_{\L}\big)^2}   \  .
   \label{tr-S2-K-mix}
\end{align}
This gives a positive contribution to the vacuum energy since $S=T-V$, which
for $\frac 1{R^2} \ll m^2_{\L}$ simplifies as
\begin{align}
 \Gamma_{\rm 1 loop}^{\cK - S^2}
 = \frac{3i}4\Tr\Big(\frac{V_{4,\cK}}{(\Box -i \varepsilon)^4}\Big)
  &\sim -\frac{1}{32\pi^2}
  \int\limits_\cM \Omega \, \r^{-2}\frac{1}{R^4}\sum_{\L;s}
  \frac{s(s+2)(2s+1) C^2_{\L}}{\mu^4_{\L}}  \
   \label{tr-S2-K-mix-simple}
\end{align}
similar to \eq{1-loop-pot-cK}, but it does not contribute to
the effective potential for $m_\cK^2$.

\subsubsection{Pure $\cK$ contribution}

Now consider the contributions arising from $V_4$
acting solely on the $6D$ terms on $\cK$. This has a similar structure as
in section \ref{sec:grav-extra-dim},
however $\Box$ acts on the non-trivial modes \eq{K-S-M-modes} with  eigenvalue
\begin{align}
 \Box = k\cdot k  + m^2_s + m^2_{\L} \ .
\end{align}
Thus the trace reduces to
\begin{align}
  \frac{3i}4\Tr\Big(\frac{V_{4,\cK}}{(\Box - i \varepsilon)^4}\Big)
  &= \frac{3i}{4(2\pi)^4}
  \int\limits_\cM d^4 x\,m_\cK^8\sum_{\L;sn} V_{4,\L} \int d^4 k\,
  \frac{1}{(k\cdot k  + m^2_s+ m^2_{\L}- i \varepsilon)^4} \nn\\
  &= \frac{3i}{4(2\pi)^4}
  \int\limits_\cM d^4 x\sqrt{G}\,m_\cK^8 \sum_{\L;sn} V_{4,\L}
  \k_{(4)}\big(m^2_s+ m^2_{\L}\big)   \nn\\
  &= -\frac{\pi^2 }{8(2\pi)^4}
  \int\limits_\cM \Omega \, \r^{-2} m_\cK^8\sum_{\L;s}
  \frac{(2s+1) V_{4,\L}}{\big(m^2_s+ m^2_{\L}\big)^2}   \  .
   \label{tr-S2-K-mix-2}
\end{align}
Again assuming $m^2_{\L}\gg \frac 1{R^2}$, this simplifies as
\begin{align}
\Gamma_{\rm 1 loop}^\cK
 = \frac{3i}4\Tr\Big(\frac{V_{4,\cK}}{(\Box - i \varepsilon)^4}\Big)
  &\sim -\frac{\pi^2 }{8(2\pi)^4}\,
  \int\limits_\cM \Omega\, \r^{-2} m_\cK^4\sum_{\L;s}
  \frac{(2s+1) V_{4,\L}}{\mu^4_{\L}}   \  .
   \label{tr-S2-K-mix-simp}
\end{align}
This is a contribution to the effective potential for $m_\cK$,
with the same structure as \eq{1-loop-pot-cK}.

\subsubsection{Pure $S^2_n$ contribution.}

Finally, we consider the pure contribution from $S^2_n$, which leads  along the lines of
\eq{S2-cov-pure} to
\begin{align}
\Gamma_{\rm 1 loop}^{S^2} =
  \frac{3i}4\Tr\Big(\frac{V_{4,S^2}}{(\Box - i \varepsilon)^4}\Big)
  &= -\frac{3}{4(2\pi)^4}
  \int\limits_\cM d^4 x\sqrt{G}\,\sum_{s,\L} \frac{\tilde V_{4,s}}{R^8}
  \k_{(4)}\big(m^2_s + m^2_{\L}\big)   \nn\\
  &= -\frac{ \pi^2 }{8(2\pi)^4}
  \int\limits_\cM \Omega\, \r^{-2} \, \sum\limits_{s,\L}\frac{\tilde V_{4,s}}{R^8}
  \frac{1}{\big(m^2_s + m^2_{\L}\big)^2}
 \end{align}
 analogous to \eq{S2-cov-pure}.
Assuming that $m^2_\L \gg \frac{1}{R^2}$, this reduces to
\begin{align}
 \Gamma_{\rm 1 loop}^{S^2,\cK}
 &\approx -\frac{\pi^2 }{8(2\pi)^4}
  \int\limits_\cM \Omega\, \r^{-2} \frac 1{m^4_\cK R^8} \sum\limits_{s,\L}
  \frac{\tilde V_{4,s}}{\mu^4_{\L}} \
  \label{tr-S2-cov-simp}
\end{align}
 which has a similar structure as \eq{S2-cov-pure}, and could a priori have either sign.

\subsection{Induced gravity in Euclidean signature}

For completeness, we also compute the induced gravity term in the case of Euclidean signature.
In the presence of fuzzy extra dimensions $\cK$, \eq{tr-M-mix} is replaced by
\begin{align}
 \Tr\Big(\frac{V_{4,{\rm mix}}}{\Box^4}\Big)
  &= \frac{2m_\cK^4}{(2\pi)^4}\int_\cM\,d^4 x\sqrt{G}\,
     \big(\sum_{\L} C^2_{\L} \tilde\k_{(4)}(m_{\L})\big) \tensor{T}{^\a^\b^{\mu}}\tensor{T}{_\a_\b^{\nu}} G_{\mu\nu} \ .
   \label{tr-M-mix-E}
\end{align}
 Here 
\begin{align}
  \tilde\k_{(4)} G_{\mu\nu}
  &= \int d^4 k\, \frac{k_\mu  k_{\nu}}{(k\cdot k + m^2)^4} \
   =  \frac{\pi^2}{12 m^2}\, \r^{-6} G_{\mu\nu} \ .
\end{align}
 Recalling that  the mass scale of the KK modes is set by the radius $m_\cK$, we obtain
\begin{align}
\boxed{\
 \Gamma_{\rm 1 loop} =
  -\frac 34 \Tr\Big(\frac{V_{4,{\rm mix}}}{\Box^4}\Big)
  = - \frac{c^2_{\cK}  }{(2\pi)^4}\int\limits_\cM d^4 x\sqrt{G}\, m_\cK^2\, \tensor{T}{_\r^\s_{\mu}}\tensor{T}{^\r_\s_{\nu}} G^{\mu\nu}  \ .
  \ }
   \label{Gamma-EH-I-euclid}
\end{align}
which has the same form as \eq{Gamma-EH-I}, with  $c_{\cK}^2$  given by \eq{C2-K-cutoff}.
This can again be written in the Einstein-Hilbert form
$\Gamma_{\rm 1 loop} \sim \int G_N^{-1}\,\cR + c\tilde T^\mu \tilde T_\mu + ... $.
Recalling our non-standard sign convention for the action,
we recover the appropriate  form $\int G_N^{-1}\cR + \cL_{\rm matter}$ of a Euclidean
gravitational action.

\subsubsection{Some technical aspects}

Before collecting the various contributions, some technical aspects should be noted.
In  the above computation, the spacetime brane $\cM$ and the
compact space $\cK$ were treated  on a different footing, using string modes on $\cM$ and harmonics on $\cK$.
It was crucial that $\cK$ is a compact fuzzy space, which admits only a finite number
of internal KK modes; otherwise the sums over these  modes would diverge.

On the other hand, it
may seem more natural to treat $\cM$ and $\cK$ on the same footing.
This is however tricky, because the product space $\cM \times \cK$ is higher-dimensional,
so that the integral over the string modes on $\cM \times \cK$ 
would no longer be almost-local. 
An appropriate  setup for such a computation
would be to use factorized quasi-coherent states $|x\rangle|\xi\rangle$
to define the string modes,
where $|\xi\rangle$ are coherent states on $\cK$.
Then the sum over the KK modes is replaced by an integral over the
 string modes on $\cK$. One can then first carry out
the integral over $\cM$, which for fixed $\xi,\xi'\in\cK$ would have precisely the
same form as e.g. in \eq{tr-M-mix}, replacing $m^2_\L$ by $|\xi-\xi'|^2$.
The (compact) integral over $\cK$ would then lead to an equivalent result as the sum over
KK modes, so that the two approaches seem perfectly cosistent.
This setup should be particularly useful if $\cK$
has a self-intersecting geometry \cite{Sperling:2018hys}; then the lowest KK modes are typically
string modes linking different sheets at their intersections, which
should provide the leading contribution to the effective potential and the
low-energy gauge theory on $\cM$.

\subsection{One-loop gravitational action and vacuum stability}
\label{sec:1-loop-grav-stab}

Combining the bare matrix model action with the one loop contributions on the space-time brane $\cM^{3,1}$ in the presence of $\cK$,
we arrive at the following one loop effective action
\begin{align}
 S_{\rm 1-loop} = S_{\rm YM} + S_{\rm grav} + S_{\rm matter}  \ .
\end{align}
Here
\begin{align}
 S_{\rm YM} = -  \frac{1}{g^2}\int\limits_\cM \Omega\, \tr_\cK\cF_{a b}\cF^{a b}
\end{align}
is the bare action of the IKKT model, while 
the induced gravitational action at one loop has the form 
\eq{tr-M-mix-2-cov}, \eq{Gamma-EH-I-cov}
\begin{align}
 S_{\rm grav}
  &= -\frac{1  }{(2\pi)^4}\int\limits_\cM d^4 x \sqrt{G}\,\r^{-2} m_\cK^2 c^2_{\cK} \,
  \tensor{T}{_\r^\s_{\mu}}\tensor{T}{^\r_\s_{\nu}} G^{\mu\nu} \
  +   \Gamma_{\rm h.o.}(\cM,\cK)
   \label{ind-grav-disc-1}
\end{align}
where the second term indicates higher-order contributions from 
the one loop effective action. The coupling to matter
-- including notably fermions\footnote{Fermions do indeed couple 
appropriately to the background geometry, as discussed in \cite{Battista:2022vvl}.}, but
also contributions from the nonabelian bosonic sector in the presence of $\cK$ 
-- is subsumed in $S_{\rm matter}$.

The most important point is the presence of an induced Einstein-Hilbert
term implicit in $S_{\rm grav}$,
with Newton constant set by the KK scale  $m_\cK^2$.
This contribution has 2 derivatives more than the bare action $S_{\rm YM}$,
and is therefore expected to dominate at shorter scales.
Since both the bare and the one loop induced action admit
 Ricci-flat linearized metric fluctuations,   the emergent gravity should be consistent with general relativity at shorter scales. 
On the other hand, the bare action $S_{\rm YM}$
is expected to govern the extreme IR regime, leading to some ''pre-gravity`` theory
at cosmic scales (cf. \cite{Sperling:2019xar,Steinacker:2019dii,Fredenhagen:2021bnw,Asano:2021phy}), where matter  no longer acts as a source of gravity.
This implies a 
cross-over behavior between these regimes, which should be studied in detail.

A detailed analysis of the resulting  gravity theory
is beyond the scope of this paper. In the following, we will
briefly study the stability of the background
with structure $\cM \times \cK$ at one loop.

\subsubsection{Effective potential and stabilization 
for basic $\cM \times \cK$ branes}

We first consider the effective potential for basic
(=non-covariant, i.e. without internal equivariant $S^2_n$ fiber) spacetime branes $\cM$
 in the presence of $\cK$ without matter.
 The semi-classical MM action contributes
\begin{align}
 S_{\rm YM} = - \int\limits_\cM \Omega\, \tr_\cK(\cF_{\dot\a \dot \b}
 \cF^{\dot\a \dot \b} +  \cF_{ij}\cF^{ij} + 2 \{T^{\dot\a},T^i\} \{T_{\dot\a},T_i\}) \
\end{align}
where the second term depends on the scale $m_\cK^2$ of $\cK$ as
in \eq{fuzzy-extra-dim-BG}
\begin{align}
 \tr_\cK(\cF_{ij}\cF^{ij}) &= m_\cK^4 F_\cK^2 \ ,
\end{align}
with  some discrete number $F_\cK^2$ is depending on the structure of $\cK$.
Note that the coupling constant $g$ is absorbed in the matrices.
We assume for now that $m_\cK$ is constant, so that $\{T^{\dot\a},T^i\} = 0$.
Together with the 
one loop contributions \eq{1-loop-pot-cK} and \eq{Gamma-EH-I} while
 dropping \eq{nonlocal-ind-grav-K} which is expected to be suppressed,
 we can  write the 
combined one loop effective  action for the background  as an effective potential
for $m_\cK$
\begin{align}
 S_{\rm 1-loop} &=: -\int\limits_\cM \Omega\, V(m_\cK)
\end{align}
as  $S = \int T - V$ in Minkowski signature.
This leads to the  quartic effective potential for $m_\cK$
\begin{align}
V(m_\cK)&= \frac{c^2_{\cK}  }{(2\pi)^4} m_\cK^2 
   \tensor{T}{^\r_\s_{\mu}}\tensor{T}{_{\r}^{\s}_{\nu}} G^{\mu\nu}
 + m_\cK^4 \Big(F^2_\cK
 + \r^{-2} \frac{\pi^2}{8(2\pi)^{4}}\sum_{\L} \frac{1}{\mu^4_{\L}} V_{4,\L}\Big) \ .
 \label{pot-1}
\end{align}
Of course
the present one-loop computation can be trusted only in the weak coupling regime, 
which means in particular that the last term (which is a 1-loop contribution) 
should be much smaller than the positive contribution $F^2_\cK$ from the bare action,
\begin{align}
 F^2_\cK \gg \r^{-2} \sum_{\L} \frac{1}{\mu^4_{\L}} V_{4,\L} \ .
 \label{weak-coupl-pot}
\end{align}
In this regime, the potential $V(m_\cK)$
has a global minimum  determined by
\begin{align}
 -c^2_{\cK} \tensor{T}{^\r_\s_{\mu}}\tensor{T}{_{\r}^{\s}_{\nu}} G^{\mu\nu}
 =  2 m_\cK^2 \Big((2\pi)^4 F^2_\cK
 + \r^{-2} \frac{\pi^2}{8}\sum_{\L} \frac{1}{\mu^4_{\L}} V_{4,\L}\Big) \ > 0
 \label{m-K-min-effpot-1}
\end{align}
provided the torsion of the background space-time  is negative in the sense that
\begin{align}
 \tensor{T}{^\r_\s_{\mu}}\tensor{T}{_{\r}^{\s}_{\nu}}G^{\mu\nu} < 0 \ .
\end{align}
This is indeed possible in Minkowski signature, but not in Euclidean signature.
Then the KK scale is determined by
\begin{align}
m_\cK^2 \approx
 - \frac{c^2_{\cK}}{2 (2\pi)^4 F^2_\cK}\, \tensor{T}{^\r_\s_{\mu}}\tensor{T}{_{\r}^{\s}_{\nu}} G^{\mu\nu} \ ,
 \label{m-K-min-effpot-2}
\end{align}
so that the induced E-H action is comparable to the geometric Yang-Mills action.
In particular, the potential \eq{pot-1} is then negative,
\begin{align}
\boxed{ \
 V(m_\cK) < 0 \ .
  \ }
\end{align}
This is an important result, which means that a background $\cM\times\cK$ with
fuzzy extra dimensions is energetically favorable compared with a background
without extra dimensions. This justifies the  presence of $\cK$,
but it remains to be clarified  which type of $\cK$ space is preferred.

\subsubsection{Effective potential and stabilization 
for covariant spacetime branes $\cM^{3,1} \times \cK$ }

We can make this more explicit for the covariant
spacetime $\cM^{3,1}$ considered in section \ref{sec:ind-grav-covar}, which describes
a FLRW space-time with metric \cite{Sperling:2019xar}
\begin{align}
 d s^2_G = -d t^2 + a^2(t)d\Sigma^2
\end{align}
with scale parameter $a(t)^{2} \sim R^2\sinh^3(\eta) \approx   \frac{3}{2}\, t$
at late times \cite{Steinacker:2020xph},
 and torsion
\begin{align}
 \tensor{T}{^\a^\b^{\mu}} = -\frac{1}{R^2}( \eta^{\a\mu} x^\b - \eta^{\b\mu} x^\a) \ .
\end{align}
Then the above 1-loop results are modified by the $\hs$ modes contributing to
the trace, as computed in \eq{tr-S2-K-mix-simp} and \eq{tr-M-mix-2-cov}.
The $S^2 - \cK$ contribution \eq{tr-S2-K-mix-simple} turns out to 
be (almost) independent of $m^2_\cK$ and will therefore be dropped here;
it does contribute to the vacuum energy as discussed below.
Assuming that $m^2_\cK \gg \frac{1}{R^2}$, the effective potential for $m_\cK$
takes again the form
\begin{align}
 V(m_\cK)&=
  \frac{c^2_{\cK}  }{(2\pi)^4} m_\cK^2\,\tensor{T}{^\r_\s_{\mu}}\tensor{T}{_{\r}^{\s}_{\nu}} G^{\mu\nu}
 + m_\cK^4 \Big(F^2_\cK
 + \r^{-2}\frac{\pi^2 }{8(2\pi)^4}\sum_{\L;s}\frac{(2s+1) V_{4,\L}}{\mu^4_{\L}}\Big) \ .
 \label{pot-cov-1}
\end{align}
Now the weak coupling condition \eq{weak-coupl-pot} is guaranteed to hold at late times, because
$\r^{2} \sim \sinh(\eta)^{3}$.
This leads again to equation \eq{m-K-min-effpot-2} for the minimum in $m_\cK$,
\begin{align}
m_\cK^2 =
 - \frac{c^2_{\cK}}{2 (2\pi)^4 F^2_\cK}\, \tensor{T}{^\r_\s_{\mu}}\tensor{T}{_{\r}^{\s}_{\nu}} G^{\mu\nu}
 \label{m-K-min-effpot-2-cov}
\end{align}
where $c_\cK^2$ is now given by \eq{C2-K-cutoff-cov}.
We can compute the rhs explicitly
using the torsion given above. This  is indeed negative, noting that
\begin{align}
 \tensor{T}{^\a^\b^{\mu}}\tensor{T}{_\a_\b^{\nu}} \g_{\mu\nu}
   = -\frac{6}{R^2} \coth^{2}(\eta) \ < 0
\end{align}
using $\g_{\mu\nu} = \sinh^{-2}(\eta)\eta_{\mu\nu}$ and $x^\mu x^\nu \eta_{\mu\nu} = -R^2\cosh^2(\eta)$ \cite{Steinacker:2020xph}.
Therefore
\begin{align}
  \tensor{T}{^\r_\s_{\mu}}\tensor{T}{_{\r}^{\s}_{\nu}} G^{\mu\nu}
 = \r^{-2} \tensor{T}{^{\dot\a}^{\dot\b}^{\mu}}\tensor{T}{_{\dot\a}_{\dot\b}^{\nu}} \g_{\mu\nu}
 = -\frac{6}{R^2\sinh^3(\eta)}
  \sim  -\frac{6}{a(t)^{2}}
\end{align}
 using \eq{torsion-contraction-frame-G}
at late times.
The minimum of the potential is thus obtained for
\begin{align}
m_\cK^2 = \frac{3 c^2_{\cK}}{(2\pi)^4 F^2_\cK }\, \frac{1}{a(t)^2}
 \label{m-K-min-3}
\end{align}
with scale set by the cosmic scale parameter, multiplied with the numerical factor
$\frac{c^2_{\cK}}{F^2_\cK} \gg 1$. This factor is very large if
$\cK = \cK_N´$ is a compact quantum space with large number of quantum cells,  because then $c^2_{\cK}$ \eq{C2-K-cutoff}
involves a sum over large representations $\L$, while $F_\cK$ does not.
We then obtain a large hierarchy between
the cosmic scale and the KK scale. The relation
between the Planck scale and the cosmic scale 
is hence given by \eq{Newton-constant-rho-mK}
\begin{align}
 \frac 1{16\pi G_N} = \frac{c^2_{\cK}}{14\pi^4}\,\r^{-2} m_\cK^2 \ \propto \
  \frac{c^4_{\cK} \r^{-2}}{F^2_\cK}\, \frac{1}{a(t)^2} \ .
\end{align}
This leads to the desired large hierarchy as long as
\begin{align}
  \frac{c^4_{\cK} \r^{-2}}{F^2_\cK}  \gg 1 \ ,
  \label{hierarchy-cond-Planck-cosm}
\end{align}
which
however decreases  with the cosmic expansion.
Then the condition $c_\cK^2 m^2_\cK \gg \frac{1}{R^2}$ \eq{K-larger-S2}
is easily met.

It is interesting to observe that $m_\cK^2$
is decreasing with the cosmic scale parameter $a(t)$.
This suggests that the Newton constant
$G_N$ is growing  with the cosmic evolution,
which is not supported by observational data.
However, the evolution of the cosmological background  and the dilaton $\r^2$
will be modified by the one loop corrections,
and further corrections are expected due to the kinetic term for $m_\cK$ discussed below.

\subsubsection{Kinetic term  and rigidity for $m_\cK$}

Since the compactification scale $m_\cK = m_\cK(x)$  is dynamical,
it is important to understand its response to deformations of the spacetime
geometry. Fortunately the kinetic term for $m^2_\cK$ implies a certain
''stiffness'',  preventing significant local variations of $m_\cK$. This
kinetic term for $m_\cK$ arises from the matrix model as
\begin{align}
 -2\int_\cM \Omega\,  \{T^\a,T^i\} \{T_\a,T_i\}
  &= -2\int_\cM \Omega\,  \{T^\a,m_\cK\} \{T_\a,m_\cK\} \cK_i \cK^i \nn\\
  &= -2\int_\cM d^4x\sqrt{G}\, (\cK_i \cK^i)\del^\mu m_\cK  \del_\mu m_\cK \ .
\end{align}
Then the equation \eq{m-K-min-effpot-2} for $m_\cK$ is replaced by
\begin{align}
-\Big(2(\cK_i \cK^i)\Box_G
+ \frac{c^2_{\cK}  }{(2\pi)^4}\,\r^{-2}\tensor{T}{^\r_\s_{\mu}}\tensor{T}{_{\r}^{\s}_{\nu}} G^{\mu\nu}
 + 2m_\cK^2 \r^{-2} F^2_\cK\Big) m_\cK = 0 \
 \label{m-K-min-effpot-3}
\end{align}
where $\Box_G$ is the d'Alembertian of the effective metric,
neglecting again the contribution from $V_{4,\L}$.
Hence $m^2_\cK(x)$ is determined such that the overall energy is minimal.
Since $m^2_\cK$ is a huge energy scale,
any significant local variation $m_\cK + \d m_\cK$ would imply a huge kinetic energy,
so that $\frac{\d m_\cK}{m_\cK} \ll 1$.
To make this more quantitative,
we observe that
\eq{m-K-min-effpot-3} multiplied with $m_\cK$ gives
\begin{align}
 m_\cK(\cK_i \cK^i)\Box_G m_\cK \sim G_N^{-1} \cR \ \sim  G_N^{-1} \Box_G\phi
\end{align}
neglecting the  $F^2_\cK$ term which is set by the cosmic background;
here $\phi$ is the Newtonian potential in linearized gravity due to some local mass distribution.
Recalling \eq{Newton-constant-rho-mK},
this leads to the estimate
\begin{align}
 \frac{\d m_\cK}{m_\cK} \ \sim \ \frac {c_\cK^2 \r^{-2}}{\cK_i \cK^i} \phi
  \ll 1 \ .
\end{align}
This is suppressed even beyond the linearized gravity regime $\phi \ll 1$
provided 
\begin{align}
 \frac {c_\cK^2 \r^{-2}}{\cK_i \cK^i} \leq 1 \ ,
\end{align}
which we shall assume. 
Then the variation $\d m_\cK$ induced by such local perturbations is strongly suppressed, so that the Newton $G_N$ is practically constant.
This condition is consistent with the condition \eq{hierarchy-cond-Planck-cosm} 
discussed in the previous section, and
 ensures that the resulting gravity
theory can be near-realistic. 
The stiffness of $m_\cK$  might also reduce the  time evolution of $m_\cK$
obtained in \eq{m-K-min-3}.

%

\subsubsection{Vacuum energy and absence of a cosmological constant}

Finally we collect the remaining terms which contribute to the potential  for the background space-time $\cM^{3,1}$. This includes
the  $S^2 - \cK$ contribution \eq{tr-S2-K-mix-simple}
and the pure $S^2$ contribution \eq{tr-S2-cov-simp}, both of which have the form
\begin{align}
 S_{\rm pot} = \int \Omega\, \r^{-2}\, V(r_\cK,R) \
 \label{vacuum-energy-1loop}
\end{align}
where
\begin{align}
\Omega =  \r_M d^4 x =  d^4x\,\sqrt{|G|} \r^{-2} \
\end{align}
is the symplectic volume form.
In orthodox quantum field theory coupled to general relativity, the vacuum energy
has the form
\begin{align}
 S_{\rm c.c.} = \int d^4 x \sqrt{|G|} \L^4 \ ,
\end{align}
which contributes to the cosmological constant $\L$.
This contribution is generically  huge in any type of path-integral approach to QFT coupled to gravity,
at least of the order of $TeV$ (which is the lowest conceivable scale of SUSY breaking),
unless some ad-hoc fine tuning is assumed.
In contrast, astronomical observations interpreted in terms of GR imply a bound on the
cosmological constant of order $meV$. This clash constitutes the cosmological constant problem, which is
 one of the most profound open problems in theoretical physics.

The present framework offers a possible resolution of this puzzle, since the quantum contributions
\eq{vacuum-energy-1loop}
to the vacuum energy are not proportional  to the
Riemannian volume form but  -- via the dilaton -- to the symplectic volume $\Omega$,
which is rigid and not dynamical.
Recall that $\Omega$ basically measures the number of states per volume, as in Bohr-Sommerfeld theory.
Hence although there is a large vacuum energy -- even for the maximally supersymmetric model,
as SUSY is spontaneously broken --  this does not gravitate as in GR,
because  it enters the action via the symplectic volume form rather than the Riemannian one.
On the other hand, the vacuum energy
will contribute to the stabilization or the choice of vacuum configuration $\cM \times \cK$
of the matrix model due to the presence of the dilaton.
It therefore appears plausible that  no cosmological constant problem arises in the present framework,
however the details of this mechanism remain to be clarified.

\section{Conclusion and outlook}

The present paper provides the explicit derivation and results for the
geometric 1-loop effective action of the IKKT matrix model on a  class of
backgrounds interpreted as noncommutative or quantized brane in target space, as announced in
\cite{Steinacker:2021yxt}.
This includes in particular the Einstein-Hilbert action, provided the background
comprises a sector describing compact fuzzy extra dimensions.
The mechanism applies at weak coupling, and is distinct from the more standard mechanism for (super-) gravity  in target space, which can be understood in the matrix model in a holographic
way at strong coupling.

The computation of the 1-loop effective action
is based heavily on a novel technique using string modes, which can be thought of as
matrix versions of open strings on the brane. This method does not rely on an expansion
into harmonics, and leads directly to the effective action in position space.

The geometric action obtained in this paper should provide the basis for a detailed investigation of the
physical properties and perspectives of the effective gravity governing the physics on the brane.
We have only briefly touched upon its numerous interesting aspects.
One issue is the selection of the background brane $\cM$, starting with its dimension;
after all, the  matrix model should provide a non-perturbative mechanism
for selecting  a vacuum. For example, the fact that the present one-loop computation
diverges on branes with dimension $\geq 8$ strongly suggests that such backgrounds are unstable;
moreover at higher loops, it is expected that only branes with dimension $4$ or less are
perturbatively stable.
This already provides a rudimentary selection mechanism, which needs to be refiend.
There are indeed on-going efforts to understand such a non-perturbative vacuum selection
using numerical methods, cf. \cite{Nishimura:2019qal,Hirasawa:2022qzg}.

Moreover, the one-loop effective potential computed in this paper for the radius or mass scale of the
compact extra-dimensional $\cK$ strongly suggest that there is  a stabilization mechanism
for $\cK$. This provides some justification for the present scenario, which clearly needs further
refinement.
In the same vein, the cosmological background $\cM^{3,1}$ will be modified by the
1-loop effective action. It is indeed remarkable that
a rather reasonable cosmological background is obtained  naturally from the classical matrix model,
without specifying any matter content. This should be refined at one loop, and related to an expected cross-over
behavior between a GR-like gravity theory valid at shorter distances, and a ``pre-gravity'' theory
arising for the classsical matrix model at cosmological scales. All these and other topics should be
addressed in further work.

\paragraph{Acknowledgments.}

I woul like to thank Emmanuele Battista for careful reading and discussions of the material.
This work was supported by the Austrian Science Fund (FWF) grant P32086.

\appendix

\section{Appendix: Some integrals}
\label{sec:some-integrals}

Consider the 4-dimensional integral
\begin{align}
 \k_{(n)} &:= \int \frac{d^4 k_\mu}{\sqrt{|G_{\mu\nu}|}}\,\frac{1}{(k\cdot k + m^2 - i \varepsilon)^n} \nn\\
 &= \r^{-4} \int \frac{d^4 k_\mu}{\sqrt{|G_{\mu\nu}|}}\,\,\frac{1}{(k_\mu k_\nu 
 G^{\mu\nu} +  m^2 - i \varepsilon)^n}  \nn\\
  &=  \r^{-4} \int d^4 k\,\frac{1}{( k_\mu k_\nu \eta^{\mu\nu} + m^2 - i \varepsilon)^n}
  \label{kappa-n}
\end{align}
where $k\cdot k = \r^{2} G^{\mu\nu}k_\mu k_\nu$ has signature $(-+++)$, and using an appropriate
redefinition of $k$ in the last step.
We can compute this by integrating first over $k^0$
via a contour rotation $\int d k^0 = i \int d k^4_E$.
Then the integral coincides with the Euclidean one up to a factor $i$, which is 
\begin{align}
  \k_{(n)} = i \r^{-4} \int d^4 k_E\,\frac{1}{(k_\mu k_\nu \eta^{\mu\nu}  + m^2)^n}
  = i 2\pi^2  \r^{-4} \int_0^\infty dk k^3 \frac{1}{(k^2 + m^2)^n} \ .
\end{align}
This is well-defined for $n\geq 3$, and gives
\begin{align}
  \k_{(4)} &= \frac{i \pi^2}{6 m^4}  \r^{-4}  \nn\\
  \k_{(3)} &= \frac{i\pi^2}{2 m^2}  \r^{-4}  \ .
  \label{kappa-3-4-explicit}
\end{align}
Note that the region which contributes to the integral is 
given by  $|k| \leq m$. This means that in the above
trace computations, $k$  has effectively  a cutoff at $m$, which should ensure that
the integrals are safely within the semi-classical regime. The size $L$
of the semi-classical wave-packets $\psi^{(L)}_{x,k}$ can then be sufficiently large, 
so that the approximation $k > \frac 1L$ \eq{k-L-regime}
easily holds for all $k$ which contribute significantly 
to these integrals. This holds provided there are no IR singularities,
which is guaranteed as long as $m^2>0$.

\bibliographystyle{JHEP}
\bibliography{papers}

\end{document}